\documentstyle[psfig,supertabular]{article}
\begin{document}
\vspace*{3.0cm}

\begin{center}
{\Large \bf Optical variability properties of high luminosity AGN
classes}
\end{center}

\vspace*{1.0cm}

{\Large \bf C. S. Stalin$^{1,2}$, Gopal-Krishna$^2$, Ram Sagar$^1$ and Paul J. Wiita$^3$} \\\\\\
\Large
$^{1}$ State Observatory, Manora Peak Nainital-263 129 \\
$^{2}$ National Centre for Radio Astrophysics, TIFR, Pune University Campus, Pune-411 007\\
$^{3}$ Department of Physics \& Astronomy, MSC 8R0314, Georgia State University, 
Atlanta, Georgia 30303-3088, USA

\vspace{5.0cm}

\noindent Address for correspondence:
\vspace*{1.0cm}

Prof. Ram Sagar

State Observatory

Manora Peak

Nainital - 263 129

Phone : (05942) 235136, 235583

Fax : (05942) 235136

Email : sagar@upso.ernet.in

~~~~~~~~~: ram\_sagar0@yahoo.co.in

\newpage

\normalsize
\begin{center}{\bf Abstract} \end{center}

\noindent We present the results of a comparative study of 
the intra-night optical variability (INOV) characteristics of 
radio-loud and radio-quiet quasars, which involves a 
systematic intra-night
optical monitoring of seven sets of high luminosity
AGNs covering the redshift range {\it z} $\simeq 0.2$ 
to {\it z} $\simeq  2.2$. The sample, matched in 
the optical luminosity -- redshift (M$_B$ -- z)  plane,
consists of seven radio-quiet quasars (RQQs), eight radio 
lobe-dominated quasars (LDQs), six radio core-dominated 
quasars (CDQs) and five BL Lac objects (BLs). Systematic
CCD observations, aided by a careful data analysis procedure,
have allowed us to detect 
INOV with amplitudes as low as 1\%.
Present observations cover a total
of 113 nights (720 hours) with only a single quasar
monitored as continuously as possible on a night.
Considering cases of only unambiguous detections
of INOV we have estimated duty cycles (DCs) of
17\%, 12\%, 20\% and 72\% respectively for
RQQs, LDQs, CDQs, and BLs. The 
low amplitude and low DC of INOV shown by RQQs compared to 
BLs can be understood in terms of their having optical
synchrotron jets which are modestly misdirected from us.
From our fairly extensive dataset, no unambiguous 
general trend of a correlation between  the INOV amplitude and 
the apparent optical brightness of the quasar is noticed.
This suggests that the physical mechanisms of INOV and long term 
optical variability (LTOV) do not
have a one-to-one relationship and different factors are
involved. Also, the absence of a clear negative correlation
between the INOV and LTOV characteristics of blazars
of our sample points towards an inconspicuous contribution
of accretion disk flucutations to the observed INOV. The INOV
duty cycle of the AGNs observed in this program 
suggests that INOV is associated predominantly with
highly polarised optical components.
We also report new VLA imaging of two RQQs (1029+329 \& 1252+020) in our
sample which has yielded a 5 GHz detection of one 
of them (1252+020; S = 1 mJy). \\

\noindent{\bf keywords:} galaxies: active --- galaxies: jets --- galaxies:
photometry --- quasars:general.

\section{Introduction}

The question of why only a small fraction of quasars are radio-loud has
been debated for almost forty years. Various arguments have been put 
forward to explain this apparent radio-loud/radio-quiet dichotomy. 
Although the reality of the dichotomy has even been questioned
(e.g., Goldschmidt et al.\ 1999; White et al.\ 2000), the most careful
analysis of the tricky selection effects indicates that it is real
(Ivezic et al.\ 2002).  It has been
argued recently that the radio emission correlates with the mass of the 
nuclear black hole (e.g., Dunlop et al.\ 2003 and references therein); however, 
this assertion has been questioned (Ho 2002; Woo \& Urry 2002). 
McLure \& Dunlop (2001) stress the importance of accretion rate and possible
changes in accretion mode to this dichotomy.

On the theoretical side, two main approaches
have been put forward to explain this
dichotomy. In one scenario, the jets in RQQs are absent or
inherently weak. Some possible mechanisms identify this
differentiating factor with the spin of the black hole (  
Wilson \& Colbert 1995, Blandford 2000), or magnetic configurations 
(Meier 2002).
At the other extreme lies the hypothesis that the relativistic
jets in RQQs are largely snuffed out before escaping the 
nuclear region itself due to heavy Inverse Compton losses 
(Kundt 2002). As a consequence, jets on radio emitting physical
scales are quenched, even though they might emit substantial amounts
of non-thermal optical synchrotron emission on micro-arc second
scales. Unfortunately, such micro-arcsec scales are beyond the 
reach of any existing imaging telescopes and the only way to probe the 
conditions at such small scales
is through flux variability observations.  

Though intra-night
optical variability (INOV) was convincingly established for 
blazars over a decade ago (Miller, Carini \& Goodrich 1989), the question 
of whether
RQQs too show INOV has remained controversial (Gopal-Krishna et al. 1995, 2000; 
Jang \& Miller 1995, 1997; Rabbette et al.\ 1998, de Diego et al. 1998; Romero et al. 1999). The cause of
INOV is still a much debated issue. However, for blazars
(CDQs and BL Lacs) which are believed to be dominated by non-thermal
Doppler boosted emission from jets (e.g., Blandford \& Rees 1978), the occurrence of rapid
intensity variations in both the radio and optical are
believed to be due to shocks propagating down their 
relativistic jets (e.g., Marscher \& Gear 1985). Intra-night variability in 
blazars may well arise from instabilities or fluctuations in the flow
of such jets (e.g., Hughes, Aller \& Aller 1992; Marscher, Gear 
\& Travis 1992). Still, alternate models, which invokes accretion 
disk instabilities or pertubations (e.g., Mangalam \& Wiita 1993; for a 
review see Wiita 1996) may also explain INOV, particularly in RQQs where
any contribution from the jets, if they are at all present, is weak.

While conclusive evidence for the presence of
jets in RQQs is far from clear, deep VLA observations hint
at the presence of weak jets even in RQQs (Kellermann et al.\ 1989; 
Miller, Rawlings \& Saunders 1993; Kellermann et al. 1994; 
Papadopoulous et al. 1995; Kukula et al.\ 1998; Blundell 
\& Beasly 1998; Blundell \& Rawlings 2001). The existence
of insipient nuclear jets in RQQs have also been inferred
from radio spectral index measurements of optically selected
quasar samples (Falcke, Patnaik \& Sherwood 1996). 
If indeed optical synchrotron jets exists even in RQQs, 
then a fairly robust signature of such un-imageable (micro-arc scale)
jets can come from detection of INOV at the level exhibited by their
radio loud counterparts, namely the LDQs. In the light of the above 
discussions, a project to search for INOV in the four major
classes of powerful AGNs was initiated in 1998 as a collaborative
effort between the State Observatory, Nainital and the National
Centre for Radio Astrophysics (NCRA), Pune. The present paper
presents many of the detailed results of this project.  Some
of these results from this large project are published 
elsewhere (Gopal-Krishna et al.\ 2003, hereafter
GSSW03; Stalin et al.\ 2003, SGSW03; Sagar et al.\ 2003, SSGW03).
 

\section[]{Selection of the sample}

AGNs in general have very
different observational characteristics including a huge range in luminosity, 
redshift and powers across the electromagnetic spectrum. Also, 
the co-moving number density of 
quasars detected at a given absolute magnitude is found to undergo 
a rapid evolution with redshift (Schmidt \& Green 1983; Boyle et al. 2000; 
Wisotzki 2000). 
Therefore, sample selection is crucial for studying INOV of quasars.
To avoid selection biases introduced by
differences in luminosity and redshift, the objects were selected such that
all objects in a given set have similar optical magnitudes, in addition to having
very similar redshifts and they are thus well matched in the optical 
luminosity--redshift plane. 
Our sample, selected from the catalog of V{\'e}ron-Cetty 
\& V{\'e}ron (1998), consists of seven sets of AGNs covering a total redshift
range from {\it z} = 0.17 to {\it z} = 2.2. Each set consists of a
radio-quiet quasar (RQQ), a radio lobe-dominated quasar (LDQ), 
a radio core-dominated quasar (CDQ)/or a BL Lac object (BL). These seven sets cover
seven narrow redshift intervals centered at {\it z} = 0.21, 0.26, 0.35, 
0.43, 0.51, 0.95 and 1.92. 

Our sample of luminous, {\it bonafide} quasars ($-$30.0 $<$ M$_B$ $<$ $-$24.3 mag),
thus overcome the selection biases such as K-correction, evolutionary effects and any
other differences introduced due to luminosity and redshift. 
Further, so as to have only {\it bona fide} RQQs in the sample, 
only those objects having a radio flux $\le$ 1 mJy at 5 GHz were selected. 

Our entire
sample consists of 26 QSOs. Of these, the number of RQQs, LDQs, CDQs (the QSO
1512+370 was initially selected as a CDQ, but it is actually a LDQ; see
SSGW03) and
BLs are 7, 8, 6 and 5 respectively; in light of the relative paucity of
BL Lacs, one object (1215$+$303) serves as a member of both Sets 1 and 2,
and none was available in the highest redshift bin.
The general properties of the objects
monitored in this program are given in Table 1. 

\section{Radio observations and reductions}
We wanted to include only those RQQs in our sample whose
fluxes available  are $\le$ 1 mJy at 5 GHz. However, 
for two of the RQQs radio fluxes were not available in the literature. 
To ascertain the radio loudness of these two QSOs 
(1252+020 and 1029+329) in 1998 we carried out observations 
using the VLA\footnote{The Very Large Array (VLA) of the 
National Radio Astronomy Observatory is operated by Associated
Universities, Inc. under a cooperative aggreement with the National 
Science Foundataion}. Snapshot obervations at 5 GHz were made using the
hybrid CnB configuration in a dual intermediate frequency 
mode with a total on-source integration time of 10 minutes.
Maps of total intensity stokes I were made from these data,
using the CLEAN algorithm available in AIPS, attaining rms
noise levels of $\sim$ 60 $\mu$Jy/beam. They are displayed in 
Fig.\ 1.  For objects with
radio detection, measurement of the flux and position 
was carried out using the AIPS task JMFIT. A quasar
is considered to have a positive detection when the radio 
source lies within 10$^{\prime\prime}$ of the optical 
position. We detected the quasar 1252+020 at
a level of 1 mJy whereas the quasar 1029+329 remained 
undetected down to a 0.2 mJy limit.

\begin{figure}
\hbox{
\psfig{file=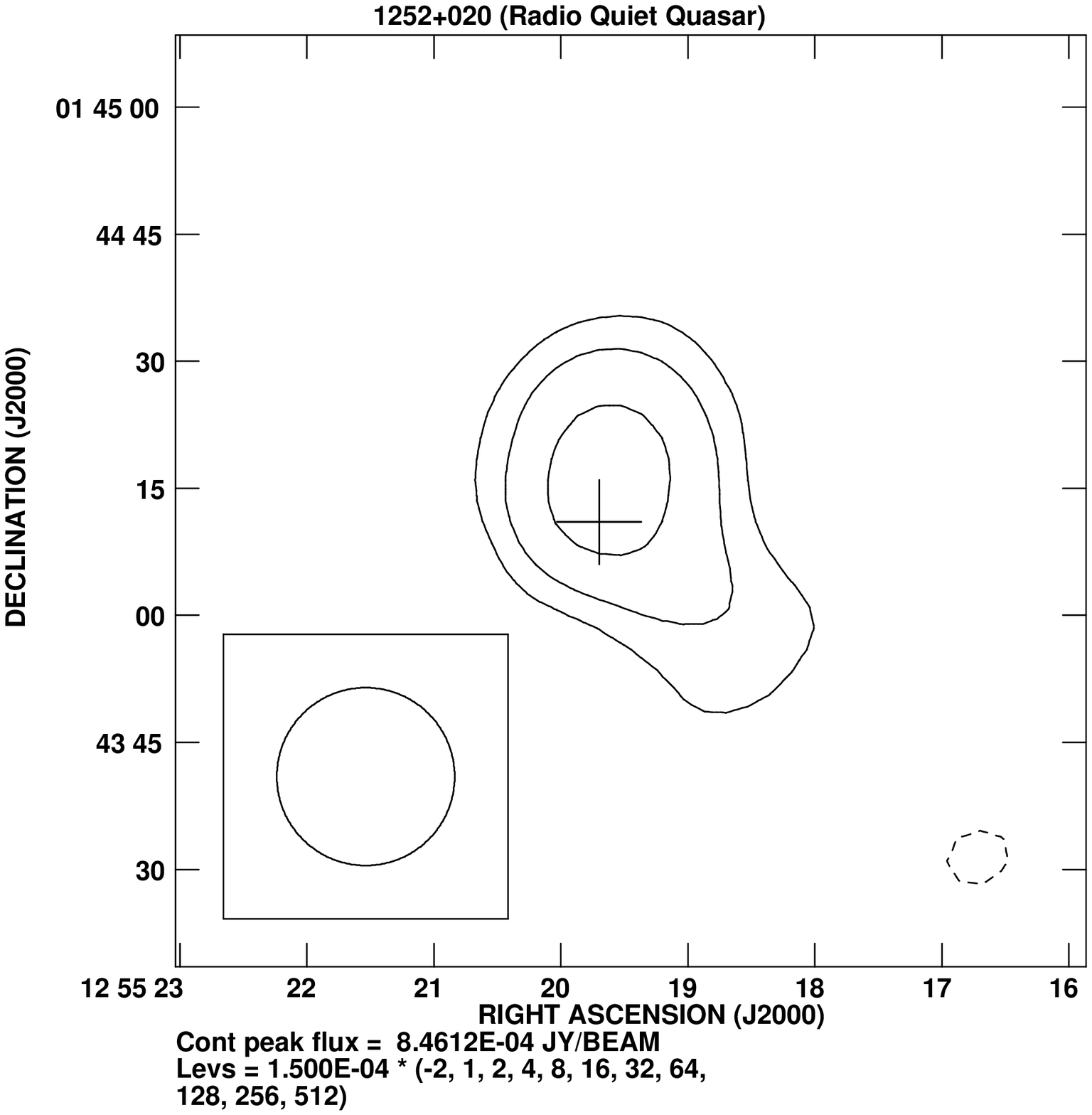,height=8cm,width=8cm}
\psfig{file=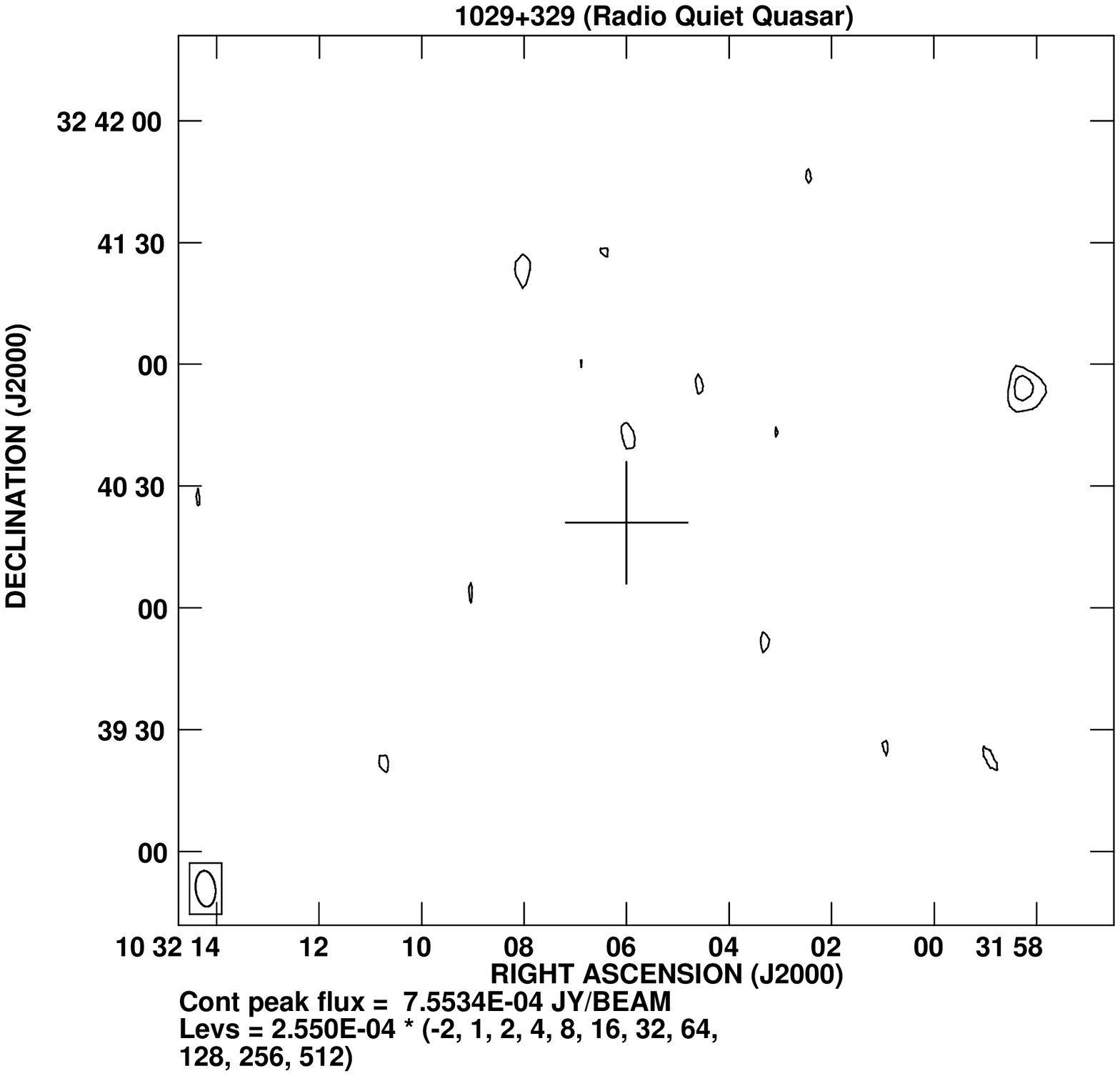,height=8cm,width=8cm,angle=-90}
}

\noindent {\bf Figure 1.} VLA 5 GHz image of the quasar 1252+020 (left) and
1029+329 (right).
\end{figure}

\section{Optical observations and Data Reductions}

\subsection{Instrument}
Observations of the quasars were carried out using the 104 cm 
Sampurnanand telescope of the State Observatory, Nainital. This is 
an RC system with a f/13 beam (Sagar 1999). The detector used for the 
observations was a cryogenically cooled 2048 $\times$ 2048 pixel$^2$
CCD, except prior to October 1999, when a smaller CCD of size
1024 $\times$ 1024 pixel$^2$ was used. In each CCD a pixel corresponds
to a square of 0.38 arcsec size, covering a total sky area of about
12$^\prime$ $\times$ 12$^\prime$ in the case of larger CCD and about
6$^\prime$ $\times$ 6$^\prime$ in the case of smaller CCD.
To increase the S/N ratio, observations were carried out in 2 $\times$ 2 binned
mode. Practically all the observations were carried out using R filter, 
except on two nights, where quasi-simultaneous R and I filter observations
were done.  The choice of R filter in this observational program is because
of it being at the maximum response of the CCD system; thus the time
resolution achievable for each object is maximised.

\subsection{Observing strategy}
The goal of this observational program was to obtain temporally dense  and 
sufficiently long duration
data trains for each object, so that one could reliably detect
statistically significant variations and determine the variability duty cycle (DC). 
Taking a clue from the optical monitoring of blazars, for which the 
probability of observing INOV in a given night is known to be greatly enhanced by  
continuous monitoring of at least 3 to 4 hours (Carini 1990), we have 
attempted to monitor each object for a minimum of 5 hours on each night.
The typical time resolution is of the order of 10 minutes, which can 
allow ultra-rapid fluctuations to be picked up. However, on some nights
time resolution is longer,  ranging up to 30 minutes for the faintest objects
in their weakest states.

Another important strategy 
adopted in this work is to sources and fields of view so as to 
ensure availablity of suitable comparison stars. The position and 
apparent R magnitudes of the comparison stars used in the 
differential photometry of our sample of quasars are
given in SGSW03 and SSGW03. Care was taken to 
have at least two, but usually more, comparison stars within 1 mag of
the QSO on the CCD frame. This allowed us to identify and discount 
any comparison star which itself varied during a given night; this
ensured reliable differential photometry of the QSO. Observations
were made on a total of 113 
for this programme during October 1998 -- May 2002.
A log of the observations, along with the results, are given in Table 2.

\subsection{Data Reduction}
Preliminary processing of the images as well as the photometry was done
using the IRAF\footnote{IRAF is distributed by the National Optical 
Astronomy Observatories, which is operated by the Association 
of Universities for Research in Astronomy, Inc. under
co-operative agreement with the National Science Foundation} software. 
The bias level of the CCD is determined from
several bias frames (generally more than seven) taken intermittently
during our observations over the night. A mean bias frame was formed using the
task {\it zerocombine} in IRAF which was then subtracted from all the image
frames of a night. Care was also taken in forming the mean bias frame such that they 
are not affected by cosmic-ray (CR) hits.  The routine step of 
dark frame subtraction was not done as the CCDs used in the 
observations were cryogenically cooled to $-$120$^\circ$ C at which the rate of
thermal charge is neglible for the exposure time of the present observations.
 Flat fielding was done by taking several 
twilight sky frames which were then median combined to generate the
flat field template which was then used to derive the final frames. 
Finally CR hits seen in the flat fielded target frames were removed using the
facilities available in MIDAS\footnote{MIDAS stands for Munich Image and
Data Analysis System and has been designed and developed by the 
European Southern Observatory (ESO) in Munich, Germany}.

\subsection{Photometry}
Aperture photometry on both the AGN and the comparison stars present on 
the flat-fielded CCD frames were carried out using the task 
{\it phot} in IRAF. A critical input to be specified to 
{\it phot} was the radius of the aperture to perform 
the photometry. The selection of this aperture determines 
the S/N for each object in the frame. In an investigation by
Howell (1989) the issue of an optimum aperture to maximize
the S/N was discussed; this optimum aperture is found close
to the FWHM of the PSF of the stars on the frame. By chosing an
aperture that is close to the FWHM of a source, clearly some of
the total flux from the source will be left out of the aperture.
However, in this work only magnitude differences are important and not the 
absolute fluxes. 
Hence, this procedure of selecting an optimum aperture
was promising for enhancing the photometric precision of the 
observations, since the S/N was maximised. The procedure we
followed for finding an optimum aperture was identical to that
used in Noble (1995). 
We  specified several
aperture radii (starting from the median FWHM of the night
and incremented by 0.2 pixels)  and then identified the aperture 
that yielded magnitudes of stars such that the scatter or variance of the 
resulting steadiest pair of Star $-$ Star differential light curves (DLCs)
 was minimum.  This process of finding
the optimum aperture is illustrated in Fig.\ 1 which shows five DLCs
for the same pair of comparison stars in the 
field of the quasar 1017$+$279 observed on 14 January 2000. The large variance
at small apertures is due to the inclusion of fewer source pixels, whereas
the large variance at large apertures is due to too many noise pixels
compared to source pixels. 

\begin{figure}
\hspace*{1.5cm}\psfig{file=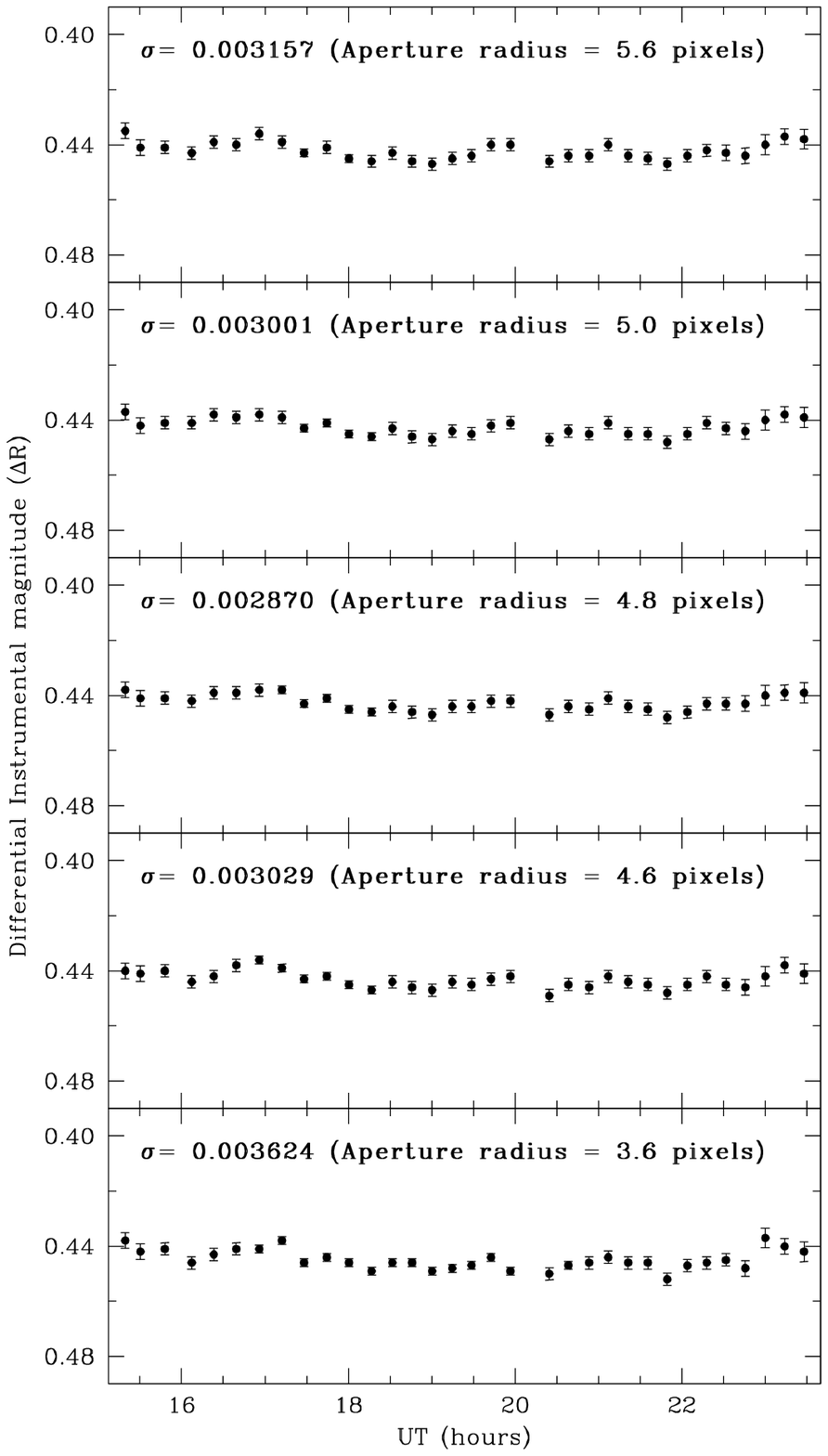,height=13cm,width=15cm}

\noindent{\bf Figure 2.} DLCs involving stars S2 and S3 in the field of the RQQ 1017+279
observed on 27 February 2000, for five different aperture radii showing 
the selection of optimum aperture. The minimum scatter occurs 
at an aperture radius of 4.8 pixels.
\end{figure}

\begin{figure}
\hspace*{2.0cm}\psfig{file=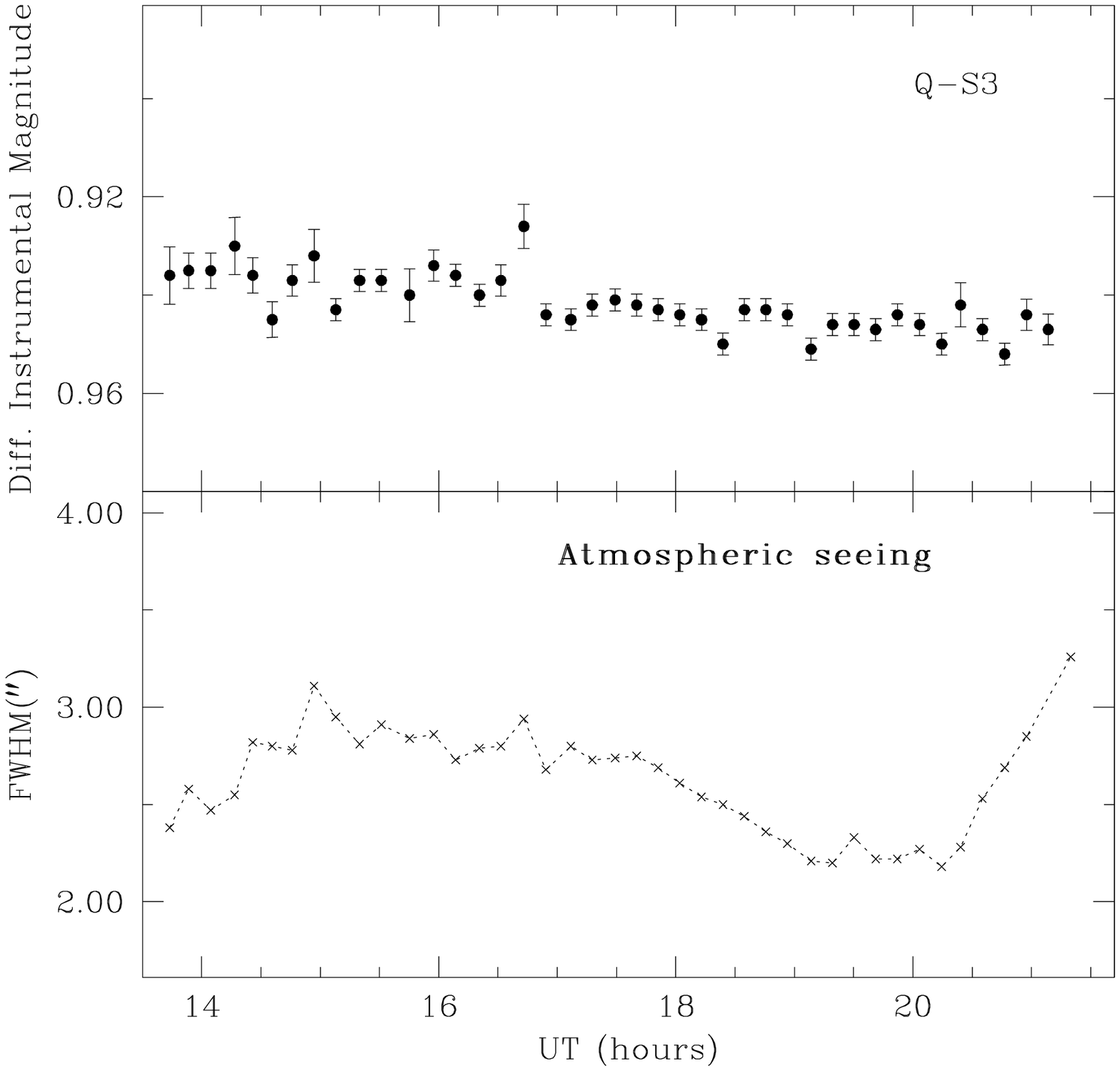,height=9cm,width=9cm}
\hspace*{-3.0cm}

\noindent {\bf Figure 3.} DLC of the LDQ 2349$-$014 observed on 17 October 2001 (top)
and the variation of seeing during that night (bottom panel). The 
aperture radius used for the photometry is 5$^{\prime\prime}$.
\end{figure}

\section{Potential sources of spurious intensity variations}

\subsection{Variable seeing}
A potential source of spurious variability in aperture photometry is 
the contamination arising from the host galaxy of the AGN recorded in the
CCD frame. This is because the surface brightness profile of any 
underlying galaxy will not respond to atmospheric seeing fluctuations in 
a manner similar to the central AGN. Thus intra-night fluctuations in the 
seeing could result in the variable contribution from the host galaxy 
within the aperture, producing spurious changes in the brightness that could
be mistaken for AGN variability. Recently the effect of spurious 
variations introudced in the DLCs by atmospheric seeing fluctuations has
been quantitatively addressed by Cellone, Romero \& Combi (2000). These 
authors conclude that spurious differential magnitude variations due to
seeing fluctuation can be substantial for AGN with brighter hosts, particularly
when small photometric apertures are used. However, our data are very
unlikely to be affected by this, since all the quasars in our sample 
are very luminous (M$_B$ $<$ $-$24.3 mag) so that the nucleus is highly 
dominant {\it vis-{\`a}-vis} the host galaxy. In fact, the underlying host
galaxy is not seen in any CCD images of our quasars except one,
(LDQ 2349$-$014) which belongs to the lowest-redshift bin of our sample.  Fig.\ 3
shows the DLC of this quasar relative to Star 3 together with
the variation of atmospheric seeing during our observing night. The lack
of any clear correlation between the quasar DLC and the seeing variations
eliminates the possibility that the steady, slow decline seen in the 
quasar DLC is an artefact arising from seeing fluctuations. The  
FWHM in our whole observing program was generally between 1.4 and 4.4 arcsec.

\subsection{Effects of colours of the comparison stars}
Even on the clearest nights, objects are dimmed due to extinction by 
the Earth's atmosphere. The amount of
dimming depends on the airmass,  the
wavelength of observation and the prevailing atmospheric conditions. The 
observed magnitude ($m_{\lambda}$) is related to the magnitude above the 
Earth atmosphere ($m_{\lambda_o}$) as (Henden \& Kaitchuck 1982)

\begin{equation}
m_{\lambda} = m_{\lambda_o} + (K_\lambda^{\prime} + K_\lambda^{\prime\prime}c)X,
\label{colour1}
\end{equation}
where $K_\lambda^{\prime}$ and $K_\lambda^{\prime\prime}$ are respectively
the  principal and second order extinction coefficients, 
$c$ is the colour index of the observed object and X is the 
airmass in the direction of the object. 
An advantage of performing differential photometry between the target and
comparison stars located on the same CCD frame is that first order extinction
effects on the differential magnitude cancels out, as both the 
comparison stars and the target are seen through nearly identical
atmospheric layers making the same X. However, $K_\lambda^{\prime\prime}$ which
applies to the colour of the object can affect the differential 
magnitude. 
From Eq.\ (1) the differential magnitude between two
objects of colour indices c1 and c2 is given by

\begin{equation}
\Delta m_{\lambda} = \Delta m_{\lambda_o} + K_\lambda^{\prime\prime}X\Delta c,
\label{colour2}
\end{equation}
where $\Delta c$ = c1 $-$ c2, is the difference between the observed colour 
indices of the  two objects. 
The relation between the standard and observed colour differences
between the two objects can be written as

\begin{equation}
\Delta C = \mu \Delta c,
\end{equation}
where $\Delta C$ = C1 $-$ C2, is the difference between the standard 
colour indices of the two objects. As $\mu$ is close to unity
(see Henden \& Kaitchuck 1982) the observed colour index difference is 
not too different from the standard
colour index difference. The values of $\Delta (B-R)$ between any 
quasar  and comparison stars in our sample range between 0.4 and $-$2.5 mag. In 
order to investigate the effects of these colour differences on the DLCs of 
the quasars under study, the following analyses have been carried out.

\begin{figure}
\hbox{
\hspace*{1.0cm}\psfig{file=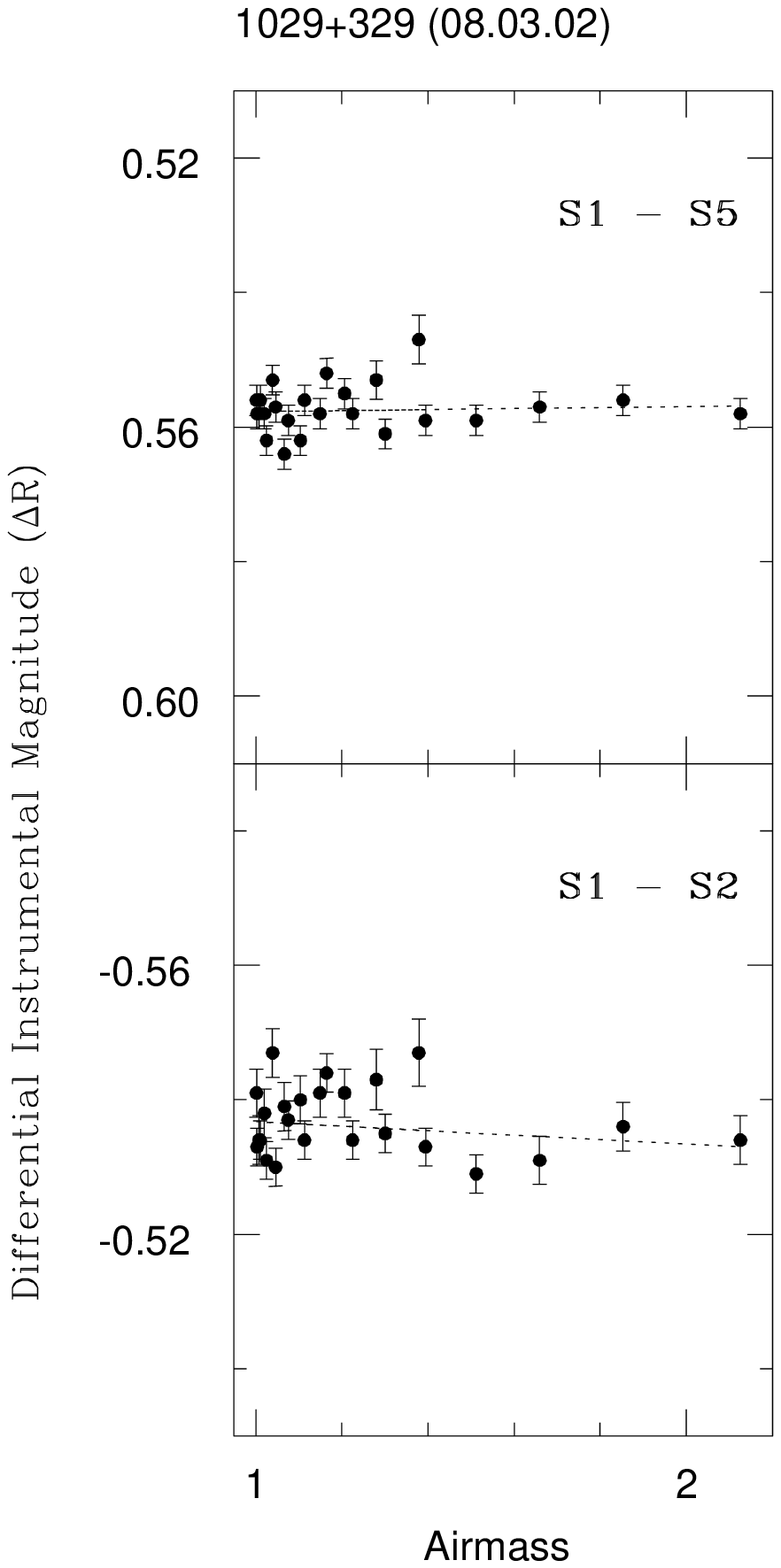,height=10cm,width=10cm}
\hspace*{-5.0cm}\psfig{file=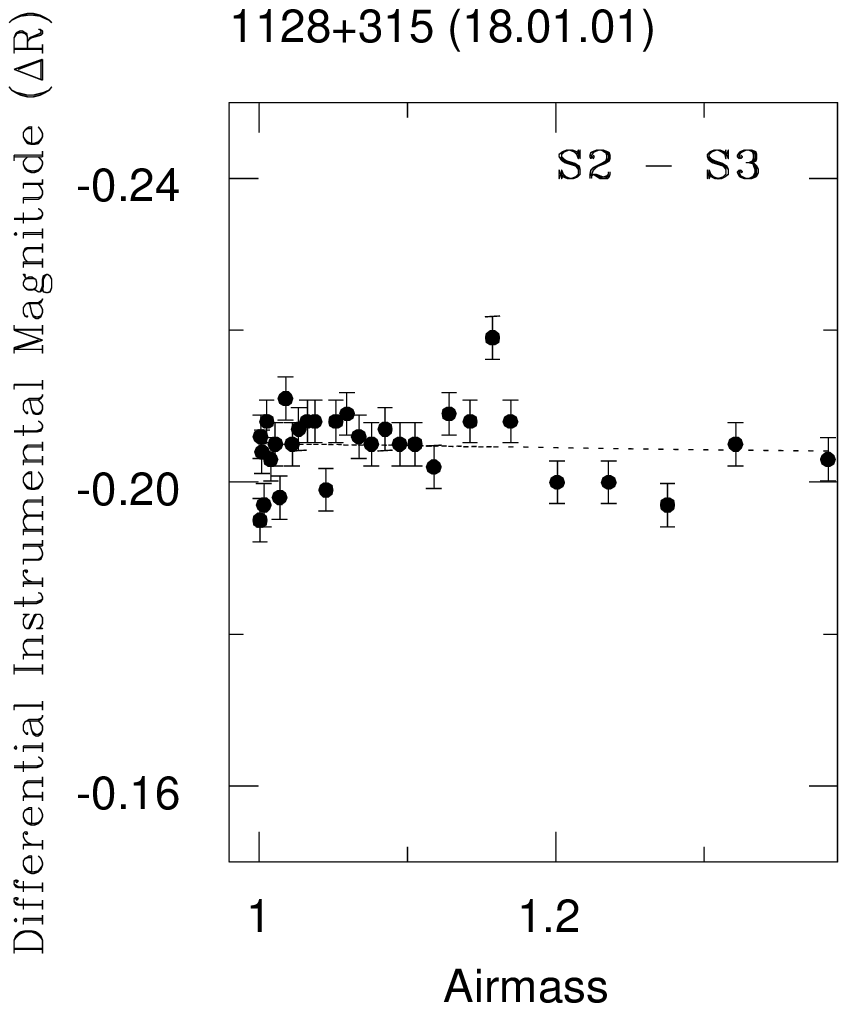,height=10cm,width=10cm}
}

\noindent {\bf Figure 4.} DLCs of the comparison stars in the field of 1029+329 (left panel)
and 1128+315 (right panel) versus airmass.
\end{figure}

From linear least square fitting of Eq.\ (2) using the standard colour index
(B$-$R) taken from the United States Naval Observatory (USNO)\footnote{http://archive.eso.org/skycat/servers/usnoa} catalog to 
the observations we found that the effect of second-order extinction 
is virtually negligible. To illustrate the effect of colour differences between
the objects on their observed DLCs, we have shown in Fig.\ 4 (top panel) the 
observed instrumental magnitude difference between the stars S1 and S5
plotted against the airmass, whereas the bottom panel shows the 
magnitude difference between S1 and S2 plotted against the airmass.
This is for the observations of the source 1029$+$329 carried out 
on the night of 8 March 2002 upto a maximum hour angle of 5 hour corresponding 
to X $\simeq$ 2.2.
Observations on other objects in our sample were also carried 
out on similar airmass at most. The colour difference 
$\Delta$(B$-$R) between S1 and S2 is 1.9 while  between S1 and S5 it is 0.9.
From Fig. 4 (left panel) it is seen that despite a rather large
range in airmass and a fairly large colour difference between
the two stars, no artificial/spurious variations are introduced.
Least square fits to both data sets shown in the left panel of
Fig. 4 yields slopes of $-$0.0008 $\pm$ 0.0023 and 0.0033 $\pm$ 0.0037,
which are essentially zero. Also shown in Fig. 4 (right panel) is 
the observed intrumental magnitude difference between stars S2 and S3, having a 
difference in colour index of $-$1.4, in the field of the QSO
1128$+$315 as observed on 18 January 2001. Linear regression 
analysis yield a slope of 0.0026 $\pm$ 0.0089, which is again 
indistinguishable from zero. This effect is thus less than the photometric
error of individual data points on the DLC and so we conclude that
the colour differences between our sets of quasars and the 
comparison stars used in the differential photometry will not
have any significant effects in the DLCs. Similar conclusions were also drawn
by Carini et al.\ (1992) in their study of rapid variability of blazars. 

\begin{figure}

\noindent {\bf Figure 5. }  Differential R-band lightcurves of radio-quiet quasars. 
The name of the object, the date and 
duration of observations are given on the top of each panel. The objects being
compared and their colour differences (in parentheses) label the right side
of each sub-panel. 
\end{figure}

\begin{figure}

\noindent {\bf Figure 5. {\it contiuned}} 
\end{figure}

\begin{figure}

\noindent {\bf Figure 5. {\it continued}} 
\end{figure}

\subsection{Error estimation of the data points in the DLCs}
Determination of the basic variability parameters, such as
peak-to-peak variability amplitudes, requires a 
realistic estimate of the photometric errors on individual
data points of a given DLC. It has been argued recently in 
the literature that the photometric errors given by the 
reduction routines in IRAF and DAOPHOT underestimate the 
true errors. Considering the difference between the 
magnitudes of the added and recovered stars using the 
{\it addstar} routine of DAOPHOT, Gopal-Krishna et al.\ (1995)
found that the formal errors returned by DAOPHOT are too
small by a factor of 1.75. Similarly, Garcia et al. (1999) found
that the error given by the {\it phot} task in IRAF is 
underestimated by a median factor of 1.73. In this work we have
 made an independent estimate of this systematic error factor.
To do this we have considered those 108 nights of 
observations (out of the 
total 113 nights) which we have found useful for INOV studies. Out of these, 
we have identified 74 DLCs  pertaining to `well 
behaved' (i.e., stable) comparison stars. Only the best available star$-$star 
DLC for each of these
nights were considered.
The unweighted mean of a DLC consisting of N data points having 
amplitude $X_i$, is given by
\begin{equation}
\langle X \rangle = \frac{1}{N}\sum_{i=1}^{N}X_i,
\end{equation}
and the variance of the DLC is

\begin{equation}
S^2 = \frac{1}{N-1} \sum_{i=1}^{N} (X_i - \langle X \rangle)^2.
\end{equation}

Both intrinsic source variability and measurement uncertainty contribute
to this observed variance. Under the assumption that both components
are normally distributed and combine in quadrature the observed
variance can be written as (see Edelson et al. 2002)

\begin{equation}
S^2 = \langle X\rangle^2\sigma_{XS}^2 + \langle\sigma_{err}^2\rangle .
\end{equation}
The first term on the right represents the intrinsic scatter induced by
source variability, and the second term is the contribution of measurement
noise as returned by the {\it phot} task in IRAF. Assuming that the 
scatter of the data points is predominantly due
to statistical uncertainty in the measurements, we have

\begin{equation}
\langle \sigma_{err}^2\rangle = \frac{1}{N}\sum_{i=1}^{N}\sigma_{err,i}^2.
\end{equation}
As we have considered here only star$-$star DLCs, the contribution of source
variability to the observed variance $S^2$ may be taken to be zero and therefore
$S^2$ becomes equal to the average of the squares of the
measurement errors $\langle \sigma_{err}^2 \rangle$.

For each DLC we computed the  quantity

\begin{equation}
Q = S^2 - \eta^2\langle \sigma_{err}^2\rangle,
\end{equation}
where $\eta^2$ is the factor by which the average of the squares of 
the measurement errors should be incremented.
For various assumed values of $\eta$ we calculated the average of Q for the
entire set of the 74 DLCs, as well as  the numbers of DLCs
for which Q was found to be positive and negative, respectively. 
The conditon of the expectation value of Q being equal to zero (i.e. $\langle Q \rangle$ = 0), 
is satisfied for $\eta$ = 1.55. On the other hand, 
the median value of Q is zero for $\eta$ = 1.40. Thus, we adopt a value of 
$\eta$ = 1.50 in further analysis. Note that this value is somewhat 
lower than 1.75 estimated in Gopal-Krishna et al.\ (1995) and  the 1.73
reported by Garcia et al.\ (1999).

\section{Optical variability}
In spite of the extensive observations (185.2 hours during 29 nights) of the 7 RQQs, 
optical variability was detected in only 3 objects during 28.1 hours on 5 nights.
Similarly, out of 8 LDQs monitored on 37 nights during 271.2 hours of observations, 
only 4 showed optical variability during 57.1 hours of 9 nights. Of the 6 CDQs, 
only 3 showed optical variability during 43.2 hours of 5 nights out of 131.2 hours
of monitoring during 20 nights. In contrast to them, all the 5 BL Lacs showed
optical variability at some times during 110.8 hours of observations on 16 nights out of the 
monitored 148 hours during 22 nights. Thus, amongst the four classes
of powerful AGNs BL Lacs are found to show maximum variability. In order
to quantify these results, we have carried out the following analysis; below
we discuss INOV and long-term optical variability (LTOV) separately. 

\subsection{Intra-night optical variability (INOV)}
In total we
have carried out 113 nights of observations, however, for 
INOV only 108 nights were considered, as the remaining 5 nights 
were too noisy to ascertain any variability (see Table 2). For blazars 
(CDQs and BL Lacs) our observations covered a total of 42 nights with an 
average of 6.6 hours/night (SSGW03). On the other hand for RQQs 
and LDQs our monitoring covered a total of 66 nights with an average of 6.1 
hours/night (SGSW03).  Out of the total of 108 observing epochs, 
nightly DLCs for 10 LDQs, 8 RQQs and 1 BL Lac and CDQ each are presented
in SGSW03, while the DLCs of another 24 nights (16 BL Lacs, 
6 CDQs \& 2 LDQs) are presented by SSGW03.
In this paper we present the INOV DLCs for the remaining 
64 nights of observation in Figs.\ 5, 6, 7 and 8 for RQQs (21 nights), 
LDQs (22 nights), CDQs (16 nights), and BL Lacs (5 nights), respectively.
This program has led to the first clear detection of
INOV in RQQs (see GSSW03). 
A clear distinction 
between the INOV nature of the two classes of 
relativistically beamed radio-loud AGNs (CDQs and BL Lacs) is found for the
first time (SSGW03). 

The variability nature of an AGN on a given night of observation is
ascertained following a statistical criterion based on the parameter 
$C_{\rm eff}$, which is similar to the parameter $C$ used by Jang 
\& Miller (1997), with the
added advantage that for each AGN we have DLCs relative to multiple comparison 
stars. The details of $C_{\rm eff}$ determination are explained in 
SGSW03. We consider a quasar to be variable if
$C_{\rm eff}> 2.57$, which corresponds to a confidence level of 
variability in excess of 99\%. A quasar is classified as a probable variable (PV)
if $2.57 \ge C_{\rm eff} > 2.00$. The values of $C_{\rm eff}$ for variable 
and probable variable quasars along with the variability status of each of the quasars monitored
in this program is given in Table 2.

We also calculated the variability amplitude of quasars which are found to be 
variable and probable variable. 
Following Romero et al. (1999), 
the amplitude of variability of a DLC is defined as 
\begin{equation}
\psi = \sqrt{(D_{max} - D_{min})^2 - 2\sigma^2},
\label{amplitude}
\end{equation}
where $D_{max}$, $D_{min}$ are the maximum and minimum in the quasar differential 
lightcurve relative to a stable star and $\sigma^2$ = $\eta^2\langle \sigma_{err}^2\rangle$.
 We have found $\eta$ = 1.50 (see Sect.\ 5.3). 
The calculated values of $\psi$ for quasars showing INOV is given in Table 2.  

\subsubsection{Structure function analysis}
Structure function analysis is a very useful pointer to the variability
characteristics of the lightcurve.
It manifests time-scales and periodicities present
in the lightcurves.  The general definition of structure functions (SFs)
and some of their properties are described, e.g., by Simonetti et al. (1985), 
Heidt \& Wagner (1996) and Hughes, Aller \& Aller (1992). 
Following Simonetti et al.\ (1985) we have
defined the first-order structure function as
\begin{equation}
D_X^1(\tau) = \frac{1}{N(\tau)}\sum_{i=1}^{N}[X(i + \tau) - X(i)]^2,
\end{equation}
where $\tau$ = time lag, $N(\tau) = \sum w(i)w(i + \tau)$ and the 
weighting factor $w(i) = 1$ if a measurement exists for the $i^{th}$ interval,
and 0 otherwise. 

The error in each point in the computed SF is

\begin{equation}
\sigma^2(\tau) = \frac{8\sigma^2_{\delta X}}{N(\tau)} D_X^1(\tau),
\end{equation}
where $\sigma^2_{\delta X}$ is the measured noise variance.

Since the sampling of our DLC is quasi-uniform, we have determined the
SF using an interpolation algorithm. For any time 
lag $\tau$, the value of $X(i+\tau)$ was calculated by linear interpolation 
between the two adjacent data points. A typical time scale in the light 
curve (i.e., time between a maximum and  a minimum, or vice versa) is 
indicated by a local maximum in the SF. In case of a monotonically
increasing structure function, the source possesses no typical time-scale
smaller than  the total duration of observations.  An indication of
periodicity can be inferred from minima of the SF.
The SF plots  as well as
their behaviour for 2 CDQs, 1 LDQ and 5 BL Lacs which showed definite 
INOV during 13 nights of observation are given in 
SSGW03. Here we present SF plots for 3 RQQs and 3 LDQs which confirmed 
INOV during 10 nights of observations in
Figs.\ 9 and 10 respectively.
The inferred variability timescale(s) and  ``period''(s) are given in Table 2. 
Because none of our light curves were long enough to show more than
two maxima or minima, it should be stressed that we are not claiming that
we have detected actual periodic components of the INOV in any of our sources.

\subsubsection{Duty cycles of intra-night optical variability (DC)}
The high precision of our data permit the estimation of INOV
duty cycle (DC) not only for different AGN classes, but also for different 
levels of variability amplitudes.  The DC of INOV of a given class
of objects is given by (Romero et al. 1999)

\begin{equation}
DC = 100 \frac{\sum_{i=1}^{n} N_i(1/\Delta t_i)}{\sum_{i=1}^{N}(1/\Delta t_i)}
  \ \     \%,
\end{equation}

\noindent where $\Delta t_i = \Delta t_{i,obs}(1 + z)^{-1}$ is the duration     
(corrected
for cosmological redshift) of a $i^{th}$ monitoring session of the source in 
the selected class. $N_i$ equals 0 or 1, depending on whether the object was 
respectively, non-variable, or variable during $\Delta t_i$. 

A DC of 17\% was found for RQQs considering only sessions for which INOV was unambiguously
detected. This value  is roughly midway between the lower values published by 
Jang \& Miller (1997) and Romero et al.\ (1999) and the higher estimate of 
de Diego et al.\ (1998). For LDQs a DC of only 12\% is found for clear detection 
of INOV; however, this rises to about 18\% if the two cases of probable detection 
are included. Also, the range of DCs for both RQQs and LDQs are found 
to be very similar (see Table 2). This similarity between LDQs and 
RQQs in terms of INOV even extends to BL Lacs if small amplitudes ($\psi < $ 3\%) 
variation is considered (GSSW03). Detailed explanations
of the similarities in the INOV characteristics of RQQs and LDQs compared
to BL Lacs are given by SSGW03.  Also, a clear distinction 
is found in DCs between the two presumably relativistically beamed AGN classes, namely 
BL Lacs and CDQs. BL Lacs are found to show a high DC, $\sim$ 72\%, whereas 
CDQs show a DC of about 20\%.  However, separating the CDQs into high polarization (CDQ-HP)
and low polarization (CDQ-LP) subsets seems to be relevant; 
CDQ-LP show very low DC ($<$ 10\%), whereas
a much higher value of DC is found for the CDQ-HP sources (51\%). It thus
appears that INOV is more closely connected to high optical polarization 
than to Doppler boosting {\it per se}.  Such 
polarized emission is commonly attributed to shocks in 
relativisitic jets (see SSGW03 for details). 

\subsubsection{Relativistic beaming, Doppler factor and accretion efficiency}
We have estimated the observed Doppler factor ($\delta_{obs}$) and accretion efficiency 
($\eta_{obs}$) in the framework of the 
relativistic beaming models (e.g., Marscher \& Scott 1980). 
The apparent 
R magnitudes of the quasars ($m_R$) were obtained from their 
observed DLCs using the apparent R magnitudes 
of the comparison stars given in USNO catalog. These 
were converted to observed monochromatic fluxes (S$_{R}$)  following 
Bessel (1979) as

\begin{equation}
S_R = 3.08 \times 10^{-23} 10^{-0.4m_R}  ~~~~~~{\rm W~m}^{-2}{\rm Hz}^{-1}.
\end{equation}

The rest-frame observed monochromatic luminosity of the source at  
frquency $\nu_{o}$ (which
we take as the frequency corresponding to the V-band) 

\begin{equation}
L_{\nu_{o}} = 4 \pi \left(\frac{cz}{H_o} \right)^2 \left(1 + \frac{z}{2} \right)^2 
S_{\nu_{o}} ~~~~~~~~{\rm for} ~~~q_o = 0,
\end{equation}
where

\begin{equation}
S_{\nu_{o}} = S_{\nu_{obs}}\left[\frac{\nu_o}{\nu_{obs}(1 + z )}\right]^{\alpha}(1 + z)^{-1},
\end{equation}
and where 
$\nu_{obs}$     = ~~frequency corresponding to R-band,
$S_{\nu_{obs}}$  = ~~flux observed in R-band, and
$\alpha       = ~~d({\rm ln} S)/d({\rm ln} \nu)$.

The observed bolometric luminosity is then calculated from this monochromatic 
luminosity, using the scaling factor given by Elvis et al.\ (1994) for 
V-band as
\begin{equation}
L_{Bol}/L_V = 13.2,
\end{equation}
where $L_V = \nu L_{\nu}$ at V-band ($\nu$ = 5.456 $\times$ 10$^{14}$ Hz)
(Elvis et al. 1994). 

We have defined  $\Delta t_{min}$ as the minimum variability
time-scale observed for a clearly detected fluctuation on a given night, corrected 
to the intrinsic value in the source frame, by dividing $\Delta t_{obs}$ with 
(1+ {\it z}).

Luminous outbursts of energy $\Delta L$ (ergs s$^{-1}$), cannot occur
on time-scales, $\Delta t_{min}$, much shorter than the light crossing time 
of the 
emitting region.  In this case, the inferred efficiency ($\eta_{obs}$) for the
conversion of accreted matter into energy for the case of spherical, 
homogeneous, non-relativistically beamed co-moving emitter is given as (Fabian \& Rees 1979)
\begin{equation}
\eta_{obs} \ge 5 \times 10^{-43}\Delta L/\Delta t_{min}.      
\label{efficiency}
\end{equation}
Assuming that the bolometric luminosity also changes in the same manner as
does the flux during outburst (e.g., Zhang, Fan \& Cheng  2002), we have estimated
$\eta_{obs}$, taking $\Delta L$ to be the variable fraction of the 
 bolometric luminosity
 during $\Delta t_{min}$.

It is known that the efficiency of energy production in the nuclear reaction
is 0.007 and that for disk accretion, the value of $\eta_{int}$ is smaller that
$\sim$ 0.3 for rapidly rotating black holes (e.g. Frank, King \& Raine 1986). 
We have calculated the lower limit of $\eta_{obs}$ 
 for all the quasars
in our sample which show definite variabilty and the results are 
given in Table 2. If $\eta_{obs}$ is found to be greater than 0.1, relativistic 
beaming is usually invoked 
to explain the observations.

Once $\eta_{obs}$ is known, one can calculate $\delta_{obs}$.
Since $\Delta L(obs) = \delta_{obs}^{3+\alpha} \Delta L(int)$ and 
$\Delta t_{min}(obs) = \delta_{obs}^{-1} \Delta t_{min}(int)$ (Worrall 1986; 
Frank, King \& Raine 1986); if we let
$\eta_{obs} = 5 \times 10^{-43}\Delta L(obs)/\Delta t_{min}(obs),$
We then find 
$\eta_{int} = 5 \times 10^{-43}\Delta L(int)/\Delta t_{min}(int),$
and thus
\begin{equation}
\delta_{obs} \ge (\eta_{obs}/\eta_{int})^{1/(4-\alpha)}.
\label{doppler}
\end{equation}

If $\eta_{int}$ is known, $\delta_{obs}$ can now be calculated.  Given that  
that $\eta_{int}$ has a range from 
0.007 for nuclear reaction to 0.32 for maximal accretion, we choose a geometric
mean for $\eta_{int}$ of $\sim$ 0.05, which is almost equal to the value of 0.057 derived 
from non-rotating accretion disk theory (e.g., Paczy{\'n}cki \& Wiita 1980) in
estimating $\delta$. The derived  
($\delta_{obs}$) for all
the variable objects with $\eta_{obs}$ $\ge$ 0.1 in our sample are given in Table  2.
The average and median values of $\eta_{obs}$ are 0.89 and 0.33 respectively, while  
$\delta_{obs}$ range between 1 to 2.55 with a mean and
median of 1.6 and 1.6 respectively.

\subsection{Long term optical variability (LTOV)}
Long term optical variability (LTOV) is seen
in 20 out of 26 objects in our sample during the period 
of our observations. The number of epochs covered range 
between three and seven and the total time span covered 
range between about a week to three years. The LTOV 
light curves for 11 quasars (6 CDQs and 5 BLs) are given by SSGW03,
while the same for another 3 quasars (2 LDQs and 1 RQQ) are presented by 
SGSW03. Here we present the 
LTOV light curves of the remaining 12 objects in our sample and comment on those
individual sources in increasing order of redshift. 

\noindent {\bf LDQ 2349$-$014}, $z = 0.174$: This QSO was
monitored for three epochs which covered a time baseline
of only five days. Over this time span the QSO was
found to vary significantly. It faded by 
about 0.03 mag over four days between 13 October 2001 and
17 October 2001 and within the next 24 hours it brightened
by about 0.05 mag (Fig.\ 11).

\noindent {\bf RQQ 0945+438}, $z = 0.226$. This object
was observed on three epochs and time baseline extends over 
nearly three years. As neither of the comparison stars 
remained stable between our first epoch and the subsequent
two epochs, we have considered only the latter
two epochs 26 February 2000 and 23 January 2001 to examine the LTOV. The
quasar dimmed by 0.07 mag within about a year between the 
two epochs (Fig.\ 12)

\noindent {\bf RQQ 0514$-$005}, $z = 0.291$. Our three epochs of
observations on this quasar covered a time baseline of 
about 10 days. The QSO did not show any change in brightness
in about 24 hours during the first two epochs, but it
faded by about 0.01 mag between 10 December 2001 and 19
December 2001 (Fig.\ 12). 

\noindent {\bf RQQ 1252+020}, $z = 0.345$: This QSO was monitored for
five epochs; however, for LTOV only four epochs could be 
considered as the first epoch (22 March 1999) lacks common 
comparison stars with the later four epochs. The total 
time baseline covered was about 2 years. The quasar
remained at the same brightness level during the first
two epochs of observations (9 March 2000 and 3 April 2000). 
It brightened by about 0.18 mag when observed a year later
on 26 April 2001. Observations on 18 March 2002 showed the
object to have dimmed by about 0.10 mag compared to 
26 April 2001 (Fig.\ 12).

\noindent {\bf LDQ 0134+329}, $z = 0.367$: Three nights of 
observations were taken on this object, but they cover a 
time baseline of just six days. The object remained
at the same brightness level during the first two days
(7 November 2001 and 8 November 2001), but dimmed
by about 0.02 mag when observed again on 13 November 2001 (Fig.\ 11)

\noindent {\bf RQQ 1101+319}, $z = 0.440$: Both brightening
and fading were clearly present in our four epochs of observations
which cover a time baseline of about two years. A fading 
of about 0.19 mag was noticed in a year between the first
two epochs of observations (12 March 1999 and 4 April 2000). 
When observed a year later (21 April 2001) a reversal
in this trend was noticed whereby the quasar brightened
by about 0.006 mag and thereafter remained steady during 
the next 24 hours (Fig.\ 12).

\noindent {\bf LDQ 0709+37}, $z = 0.487$: The LTOV nature of 
this QSO can be gleaned from the five epochs of 
observations done in the year 2001. It remained at 
the same brightness level during the first three epochs
of observations during January 2001. However, it 
brightened by about 0.03 mag when observed on 
20 December 2001 and then faded by about 0.09 mag
within the next 24 hours (between 20 and 
21 December 2001; see Fig.\ 11).

\noindent {\bf RQQ 1029+329}, $z = 0.560$: This QSO was
monitored on six nights; however, for characterizing the
LTOV only the later five epochs were considered due to 
the lack of common comparison stars in the first epoch
of observation. The QSO remained at the same brightness
level during our last five epochs of observations 
(between 2 March 2000 and 8 March 2002) covering
a two year baseline (Fig.\ 12)

\noindent {\bf LDQ 0350$-$073}, $z = 0.962$: The quasar was 
found not to show any variability during our three epochs
of observations within a week (during 14 
to 18 November 2001) (Fig.\ 11).

\noindent {\bf LDQ 0012+305}, $z = 1.619$. The quasar
was found to be variable from our six epochs of observations
encompassing about 10 months during 2001. It remained
at the same brightness level during the first three epochs of
observations and then faded by about 0.06 mag when observed
nine months later on 14 October 2001. It then remained at the 
same brightness level for the next one week during which it
was monitored on three nights (Fig. 11)

\noindent {\bf RQQ 1017+279}, $z = 1.918$. This quasar was
monitored on three epochs. However, its LTOV could only be 
probed from the later two epochs, due to the lack of common 
comparison stars between them and the first epoch. The quasar
remained at the same flux level between the last two epochs
(i.e., 14 January 2000 and 27 February 2000) separated by 
seven weeks (Fig. 12)

\noindent {\bf CDQ 1225+317}, $z = 2.219$. No evidence of variability
is found over the 2 year time (between March 1999 and April 2001) 
covered by our three epochs of observations (Fig. 11).

\section{Do INOV and LTOV have a common origin?}
A perception has gradually been developed that the most common cause of rapid 
optical variability of blazars is associated with disturbance
occuring within their relativistic jets. For instance, Romero et al. (1999)
have argued that while the LTOV arises due to large-scale relativistic 
shocks propogating through the jet, the INOV occurs due to the interception
of these shocks by small scale magnetic irregularities in the jet. To 
verify these ideas, one needs to have a really large database of INOV monitoring observations
at multiple epochs. Nonetheless, our fairly extensive database allows us to 
perform a check by searching for a correlation between the INOV amplitude
and the level of optical brightness of the quasar. To do this, we have
considered a subset of 5 `blazars' in our sample,  
for which INOV was clearly detected on at least two nights. The LTOV
information was obtained from our multi-epoch data. This was done
by further identifying at least one stable (`well behaved') comparison
star common to all the epochs and then calculating the mean optical magnitude
of the quasar relative to that star for each epoch. 
The night for which the quasar had the minimum 
optical brightness is taken as the base level for LTOV and the resulting offset values
of the quasars magnitudes are given in Table 3, in the sequence of 
increasing brightness. Thus, these data provide a decent quantitative 
description of the LTOV of the quasars. It may, however, be noted that due 
to the scheduling constraints, the total time span covered is highly variable
from quasar to quasar, as can be seen from our log of observations in Table 2. 

Fig.\ 13 shows the comparison
of LTOV and INOV amplitudes for the subset of 5 blazars and one LDQ following the 
prescription above. 
Different symbols are used for each
object and the multi-epoch data for a given object are joined by  
dashed lines. It is fair to state that, based on our dataset, no
unambiguous general trend for a correlation between the INOV and LTOV 
amplitudes is evident for these luminous AGNs. Thus, from an observational point of view, 
there is as yet no strong case for assuming common excitation conditions
for LTOV and INOV in the optical output of blazars.

\subsection{Do accretion disks contribute to INOV?}
The relation of INOV to the long term variablity nature
of quasars can be used to ascertain the contribution of disk emission 
to INOV  (e.g., Mangalam \& Wiita 1993). 
Several Optically Violent Variables (OVVs) including 3C 345 (Bregman et al.\ 1986; Smith et al.\ 1986),
PKS 0420$-$014, B2 1156$+$295 and 3C 454.3 (Smith et al. 1988) supply
photometric and polarimetric evidence for substantial accretion
disk contributions to their emission. But the general absence of 
a Big Blue Bump in BL Lac
spectra does not provide the strong evidence in favour of an accretion
disk components that exists in most quasars (Sun \& Malkan 1989).
Still it has been claimed that in one of the best studied BL Lacs,
2155$-$304, an accretion disk does seem to contribute to their spectrum
(Wandel \& Urry 1991). However, optical polarimetry of the source 
implies that the disk contribution, if present, is fairly small
(Smith \& Sitko 1991). 

If INOV is found to be inversely 
correlated to the brightness state of the object this may provide
evidence for the accretion disk contribution to the observed INOV. 
This is because the disk component will be relatively more important when
the object is in a quiescent state and jets are weaker. 
A comparison of INOV
against LTOV for all the BL Lacs in our sample is shown in Fig.\ 13.
No correlation between INOV and LTOV is noticed suggesting little
contribution of an accretion disk to the observed INOV in any of 
these sources.

\subsection{INOV in the context of the starburst models of AGN}
In the starburst model (Terlevich et al. 1992; Terlevich \& Melnick 1985) of AGNs, variability 
results from random
superposition of events such as supernova explosions generating rapidly
evolving compact supernova remnants (cSNRs) due to the interaction
of their ejecta with the high density circumstellar environment. This
model is supported by the striking similarity between the optical spectra
of some AGNs and of cSNRs (Filippenko et al.\ 1989). The characterisitics of 
an event (i.e., light curve, amplitude and time-scale) results from
the combination of complicated processes (Terlevich et al.\ 1992).
Still, the lightcurves of AGNs of  various absolute 
luminosities and redshifts can be predicted from the model and are 
found to be reasonably consistent with the observed dependence of structure function 
(the curve of growth of variablity with time) on 
luminosity and redshift (cid Fernandes, Aretxaga \& Terlevich 1996; Cristiani et al. 1996).
The lightcurves of cSNRs are still poorly known.  
While they seem to be  
consistent with the optical lightcurves of the low luminosity AGNs, 
NGC 4151 and NGC 5548 (Aretxaga \& Terlevich 1994),  
as argued by cid Fernandes, Aretxaga \& Terlevich (1997) and references therein,  
starburst models may be
useful in explaining variability of modest amplitude in very weak AGN
($M_B \sim -20$ is the peak of a cSNR).  As the energies involved
in the intra-night fluctuations we have detected are at least an order of magnitude
above these, it
appears that such a model has negligible applicability to the variability 
events associated even  with the non-blazar type AGNs (Tables 2 and 3). 

\subsection{INOV and polarization: instances of exceptional behaviour}
High and variable optical polarization is one of the 
defining characterisitc of blazars (Angel \& Stockman 1980)
whose radiation is now believed to be dominated by relativistically 
boosted non-thermal jet emission (Blandford \& Rees 1978). To 
investigate the relationship of optical INOV to the 
polarization,  we have plotted the duty cycles of the 10 
variable quasars in our sample against their 
optical polarizations (Fig.\ 14). It is found that
the INOV is seen more in high polarization quasars ($P_{opt}$ $>$ 3\%).
The main exception in our sample is, curiously,  the BL Lac object 
0735+178 for which a very high degree of optical polarization ($P_{opt}$ = 14.1\%) 
has been reported (Table 1). Moreover, a CDQ, 1308+326, even though highly
polarized ($P_{opt}$ = 10.2\%) has shown INOV on just one of the five
nights on which it was monitored (for 6 to 8 hours duration each time).
However, in this case,  the non-detection of INOV on four of the nights could well be
the result of the blazar being too faint, leading
to abnormally noisy DLCs which would mask more typical variations. 
In contrast, an exceptionally noteworthy case is the 
LDQ 2349$-$014 which is very weakly polarized ($P_{opt}$ = 0.91\%) and yet
 showed INOV on all the 
three nights it was monitored. However  
its INOV is of low level ($\psi$ $<$ 3\%) and may even be
the part of a longer time-scale variation. The weak correlation visible
in Fig.\ 14, along with our results of 
the DC of INOV discussed above (Sect.\ 6.1.2) argues that INOV is closely 
associated with optical polarization.

\section{Conclusions}
In this extensive optical monitoring programme, an effort has been made to understand 
the INOV characteristics of the four major classes of luminous AGNs, namely
radio-quiet quasars, radio-lobe dominated quasar, radio-core dominated quasars and
BL Lac objects. The major finding of this present study are:

\begin{enumerate}
\item The first clear evidence of INOV in RQQs has been found from this
observational programme (Gopal-Krishna et al. 2003).
\item BL Lac objects are found to show high DC of INOV (72\%).
In contrast, LDQs, CDQs and RQQs show much smaller INOV duty cycles
($\sim$20\%).  Although our estimate of the DC of
RQQs is much higher than 3\% found by Romero et al. (1999),
it clearly falls much short of the DC for BL Lacs. In this sense,
there is still a marked
difference between the INOV properties of radio-loud and
radio-quiet quasars, in contrast to the conclusions reached by de Diego et al. (1998),
namely that INOV occurs as frequently in RQQs as it does in RLQs. This
different conclusion of ours is probably due to our higher sensitivity,
and the distinctly temporally denser monitoring
achieved in our program.

\item The high DC shown by BL Lacs strongly suggests that
relativistic beaming plays an important role in their observed
micro-variability. Nonetheless,
the lower DC and amplitude of INOV, shown by LDQs  and RQQs can be
understood within the framework
of unified models,  as these objects are believed to have modestly
misaligned jets (see GSSW03). We infer that LDQs and even RQQs
could  posses relativistic, optical synchrotron nuclear
jets on micro-arcsecond scales. Their intrinsic variabilty can thus be similar
to BL Lacs, and their observed much milder INOV can be attributed to
their (micro) jets being modestly misaligned from our
direction (and so  having lower Doppler factors $\delta$).
Thus, the observed difference in the
micro-variability nature of LDQs and RQQs compared to BL Lacs can be
accounted for in terms of their optically
emitting nuclear jets undergoing different degrees of
Doppler boosting in our direction. Observers located in suitably
different directions may well find these same LDQs and RQQs to be
large-amplitude rapid variables (GSSW03).

\item For blazars, no correlation is noticed between their INOV
amplitude and theie apparent optical brightness (Fig.\ 13). This suggests that
the physical mechanisms of intra-night and long-term optical variability
do not have one-to-one relationship, and other factors are involved.
Likewise, the absence of a clear negative correlation between the INOV and LTOV
characterisitics of the blazars in our sample points towards
an inconspicuous contribution of accretion disk to the observed 
INOV, though we stress that our sample size is very small and no
firm conclusions can yet be drawn.

\item A clear distinction is found for the first time between the
INOV properties of the two classes of relativistically
beamed radio-loud AGNs (RLQs), namely, BL Lacs and CDQs.
The latter are found to exhibit low INOV 
duty cycle, no more than that exhibited by RQQs and LDQs.
Moreover BL Lacs show high amplitude and DC of INOV.  But in confining the
discussion only to 
the CDQs which have high optical polarization, it is found that they 
resemble BL Lacs, both in amplitude and DC of INOV. It thus appears
that the mere presence of a prominent (and hence Doppler boosted) radio
core does not guarantee INOV; rather it appears that the more crucial 
factor is the optical polarization of the core emission. Such polarized
emission is normally associated with shocks in a relativistic jet. 
This strongly suggest that the INOV
is associated  predominantly with highly polarized
quasars (see SSGW03).

\item Even though the percentage luminosity variations implied by 
the INOV for these luminous AGNs is small, the total power involved
is still so enormous so as to render a starburst/supernova
explanation untenable for these rapid events.
\end{enumerate}

\section*{Acknowledgments}

The help rendered by the technical staff at the 104 cm telescope of State Observatory, 
Nainital is thankfully acknowledged.
This research has made use of the NASA/IPAC Extragalactic Database (NED), which is operated
by the Jet Propulsion Laboratory, California Institue of Technology, under contract with the 
National Aeronautics and Space Administration. CSS thanks NCRA for hospitality and use of
its facilities. PJW's efforts were partially supported by Research Program Enhancement funds
at GSU, and he is grateful for hospitality at Princeton University Observatory.


\newpage


\begin{table*}
\caption{General information on radio-quiet, lobe-dominated, 
core-dominated and BL Lac objects monitored in the 
present programme. The quasar 1512+370 initially selected as a CDQ is actually a LDQ (see SSGW03)}
\vspace*{0.5cm}

\begin{tabular}{llllllllrlr} \hline \hline
Set     &Object & Other Name  & Type & RA(2000)  & Dec(2000)   &  ~~B    & ~~M$_B$  & {\it z} & \%Pol$^\dag$ & log R$^{\ddag}$ \\
No.     &       &             &      &           &             & (mag)   & (mag)    &           &  (opt)       &           \\ \hline
1.      &0945+438   & US 995          & RQQ  &09 48 59.4 &   +43 35 18 & 16.45  & $-$24.3  & 0.226  &  ~~~---   & $< -$0.07  \\       
        &2349$-$014 & PKS 2349-01     & LDQ  &23 51 56.1 & $-$01 09 13 & 15.45  & $-$24.7  & 0.174  & 0.91      &  2.47     \\
        & 1309+355  & PG 1309+355     & CDQ  &13 12 17.7 &   +35 15 23 & 15.60  & $-$24.7  & 0.184  & 0.31*   &  1.36    \\
        & 1215+303  & B2 1215+30      & BL   &12 17 52.0 &   +30 07 01 & 16.07  & $-$24.8  & 0.237  & 8.0  &  2.63    \\
2.      &0514$-$005 & 1E 0514-0030    & RQQ  &05 16 33.5 & $-$00 27 14 & 16.26  & $-$25.1  & 0.291  & ~~~---    & $<$ 0.06  \\
        &1004+130   & PG  1004+130    & LDQ  &10 07 26.2 &   +12 48 56 & 15.28  & $-$25.6  & 0.240  & 0.78      &  2.29     \\
        & 1128+315  & B2 1128+31      & CDQ  &11 31 09.4 &   +31 14 07 & 16.00  & $-$25.3  & 0.289  & 0.62   &  2.43    \\
3.      &1252+020   & Q  1252+0200    & RQQ  &12 55 19.7 &   +01 44 13 & 15.48  & $-$26.2  & 0.345  & ~~~---    &  $-$0.28    \\
        &0134+329   & 3C 48.0         & LDQ  &01 37 41.3 &   +33 09 35 & 16.62  & $-$25.2  & 0.367  & 1.41      &  3.93    \\
        &1512+370   & B2 1512+37      & LDQ  &15 14 43.0 &   +36 50 50 & 16.25  & $-$25.6  & 0.370  & 1.17   &  3.57    \\
        & 0851+202  & OJ 287          & BL   &08 54 48.8 &   +20 06 30 & 15.91  & $-$25.5  & 0.306  &  12.50  &  3.32    \\
4.      &1101+319   & TON 52          & RQQ  &11 04 07.0 &   +31 41 11 & 16.00  & $-$26.2  & 0.440  &  ~~~---   &$<-$0.41  \\
        &1103$-$006 & PKS 1103$-$006  & LDQ  &11 06 31.8 & $-$00 52 53 & 16.39  & $-$25.7  & 0.426  & 0.37      &  2.80     \\
        &1216$-$010 & PKS 1216$-$010  & CDQ  &12 18 35.0 & $-$01 19 54 & 16.17  & $-$25.9  & 0.415  & 6.90**  &  2.34    \\
        &0735+178   & PKS 0735+17     & BL   &07 38 07.4 &   +17 42 19 & 16.76  & $-$25.4  &$>$0.424& 14.10  &  3.55    \\
5.      &1029+329   & CSO 50          & RQQ  &10 32 06.0 &   +32 40 21 & 16.00  & $-$26.7  & 0.560  & ~~~---    &$<-$0.64  \\
        &0709+370   & B2 0709+37      & LDQ  &07 13 09.4 &   +36 56 07 & 15.66  & $-$26.8  & 0.487  & ~~~---    & 2.08      \\
        &0955+326   & 3C 232          & CDQ  &09 58 20.9 &   +32 24 02 & 15.88  & $-$26.7  & 0.530  & 0.53  & 2.74     \\
        &0219+428   & 3C 66A          & BL   &02 22 39.6 &   +43 02 08 & 15.71  & $-$26.5  & 0.444  & 11.70  & 2.83     \\
6.      &0748+294   & QJ 0751+2919    & RQQ  &07 51 12.3 &   +29 19 38 & 15.00  & $-$29.0  & 0.910  & ~~~---    & $-$0.68     \\
        &0350$-$073 & 3C 94           & LDQ  &03 52 30.6 & $-$07 11 02 & 16.93  & $-$27.2  & 0.962  & 1.42      & 3.07     \\
        &1308+326   & B2 1308+32      & CDQ  &13 10 28.7 &   +32 20 44 & 15.61  & $-$28.6  & 0.997  & 10.20  & 2.71     \\
        &0235+164   & AO 0235+164     & BL   &02 38 38.9 &   +16 37 00 & 16.46  & $-$27.6  & 0.940  &  14.90   & 3.29     \\
7.      &1017+279   & TON 34          & RQQ  &10 19 56.6 &   +27 44 02 & 16.06  & $-$29.8  & 1.918  & ~~~---    &$<-$0.49  \\
        &0012+305   & B2 0012+30      & LDQ  &00 15 35.9 &   +30 52 30 & 16.30  & $-$29.1  & 1.619  & ~~~---    & 1.76     \\
        &1225+317   & B2 1225+31      & CDQ  &12 28 24.8 &   +31 28 38 & 16.15  & $-$30.0  & 2.219  &  ~~0.16   & 2.26     \\
\hline\hline
\end{tabular}

\hspace*{-0.1cm} $^\ddag$ log R is the ratio of the radio-to-optical flux densities  calculated following Stocke et al. (1992)

\hspace*{-0.1cm} $^\dag$ Optical polarization are from Wills et al. (1992) except those marked with $\ast$ (Berriman et al. 1990)
                    and $\ast\ast$ (Hutsemekers \& Lamy 2001)
\end{table*}

\newpage

\begin{table*}
\noindent {\bf Table 2.} Results of INOV observations of QSOs. N is the total number of points, and T is the 
duration, of each observation. Status of INOV are denoted as V (for variables), PV (for probable variables) and 
NV (for non-variables). $C_{eff}$ and $\psi$ indicate the statistical significance and amplitude of variability.
The variability timescale and periods are denoted as $\tau$ and $P$ respectively. $\eta_{obs}$ and $\delta_{obs}$
denote the accretion efficiency and doppler factor in the frame work of relativistic beaming models. The last
column gives the references where DLCs are shown.

\vspace*{0.5cm}

\small
\begin{tabular}{llllcclcccccl} \hline
Set & Object    &  Date     & N    & T & INOV  & C$_{eff}$ & $\psi$ & $\tau$ & P & $\eta_{obs}$ & $\delta_{obs}$ & Ref. to  \\
No. &           &           &      & (hr)  & Status$^*$&           &   \%     &  (hr)      & (hr) &      &         &   DLCs   \\ \hline
1.  & 0945+438  &  15.01.99 &  44  & 8.0 & NV  &        &       &          &             &       &      & Present work \\
    & (RQQ)     &  26.02.00 &  31  & 6.3 & NV  &        &       &          &             &       &      & Present work\\
    &           &  23.01.01 &  24  & 6.6 & NV  &        &       &          &             &       &      & Present work\\
    & 2349-01   &  13.10.01 &  34  & 6.8 & V   & 3.6    &  2.2  &          &             &  0.02 &      & SGSW03\\
    &  (LDQ)    &  17.10.01 &  39  & 7.6 & V   & 3.1    &  1.5  &          &             &  0.02 &      & SGSW03\\
    &           &  18.10.01 &  40  & 7.7 & V   & 3.2    &  1.6  &  5.0     &             &  0.02 &      & SGSW03\\
    & 1309+355   & 25.03.99 &  39  & 6.7 & NV  &        &       &          &             &       &      & Present work\\
    &  (CDQ)     & 01.04.01 &  32  & 4.6 & NV  &        &       &          &             &       &      & Present work\\
    &            & 02.04.01 &  41  & 5.2 & NV  &        &       &          &             &       &      & Present work\\
    & 1215+303   & 20.03.99 &  21  & 7.0 & V   & 5.5    & 3.5   &  4.2     &             &  0.74 & 1.8  & SSGW03\\
    &(BL)        &25.02.00 &  28  & 5.9 & NV  &        &       &          &             &       &      & Present Work\\
    &            & 31.03.00 &  27  & 5.0 & NV  &        &       &          &             &       &      & Present Work\\
    &            & 19.04.02 &  23  & 6.8 & V   & 4.9    & 1.8   &  $>$ 6.8 &             &  0.05 & 1.0  & SSGW03\\
2.  & 0514$-$005&  09.12.01 &  25  & 5.3 & NV  &        &       &          &             &       &      & Present work\\
    &  (RQQ)    &  10.12.01 &  23  & 5.8 & NV  &        &       &          &             &       &      & Present work\\
    &           &  19.12.01 &  35  & 7.5 & NV  &        &       &          &             &       &      & Present work\\
    & 1004+130  &  27.02.99 &  30  & 4.3 & NV  &        &       &          &             &       &      & Present work\\
    &   (LDQ)   &  16.02.99 &  36  & 6.5 & NV  &        &       &          &             &       &      & Present work\\
    &           &  29.03.00 &  21  & 3.8 & NV  &        &       &          &             &       &      & Present work\\
    &           &  30.03.00 &  26  & 4.6 & NV  &        &       &          &             &       &      & Present work\\
    &           &  18.02.01 &  42  & 5.5 & NV  &        &       &          &             &       &      & Present work\\
    &           &  24.03.01 &  50  & 6.4 & NV  &        &       &          &             &       &      & SGSW03 \\
    & 1128+315   & 18.01.01 &  31  & 5.7 & NV  &        &       &          &             &       &      & Present work\\
    & (CDQ)      & 09.03.02 &  27  & 8.2 & NV  &        &       &          &             &       &      & SGSW03 \\
    &            & 10.03.02 &  28  & 8.3 & NV  &        &       &          &             &       &      & Present work\\
3.  &  1252+020 &  22.03.99 &  36  & 6.4 & V   & 3.3    &  2.3  &          &             &       &      & SGSW03\\
    &   (RQQ)   &  09.03.00 &  29  & 6.1 & NV  &        &       &          &             &       &      & Present work \\
    &           &  03.04.00 &  19  & 4.3 & V   & 3.6    &  0.9  & $>$4.3   &             &  0.04 &      & SGSW03\\
    &           &  26.04.01 &  20  & 4.6 & NV  &        &       &          &             &       &      & Present work \\
    &           &  18.03.02 &  19  & 7.3 & NV  &        &       &          &             &       &      & Present work \\
    &  0134+329 &  07.11.01 &  33  & 6.5 & NV  &        &       &          &             &       &      & Present work \\
    &   (LDQ)   &  08.11.01 &  32  & 6.7 & NV  &        &       &          &             &       &      & Present work \\
    &           &  13.11.01 &  46  & 8.6 & NV  &        &       &          &             &       &      & Present work \\
    & 1512+370   & 23.03.02 &  24  & 7.0 & NV  &        &       &          &             &       &      & Present work\\
    & (LDQ)      & 27.03.02 & 28&7.0& NV  &        &       &          &             &       &      & Present work\\
    &            & 21.04.02 &  11  & 4.3 & V   & 2.8    & 2.6   &  0.4     &             & 0.53  & 1.60 & SSGW03\\
    &            & 23.04.02 &  15  & 5.3 & NV  &        &       &          &             &       &      & Present work\\
    &            & 01.05.02 &  19  & 6.6 & V   & 3.0    & 3.9   &  0.4     & 1.8,3.4,5.6 & 0.18  & 1.29 & SSGW03\\
    & 0851+202   & 29.12.98 &  19  & 6.8 & V   & 2.8    & 2.3   & $>$ 6.8  &             & 0.12  & 1.20 & SSGW03\\
    & (BL)       & 31.12.99 &  29  & 5.6 & V   & 6.5    & 3.8   &  3.0     &             & 0.07  & 1.07 & SSGW03\\
    &            & 28.03.00 &  22  & 4.2 & V   & 5.8    & 5.0   &  1.2     &             & 0.88  & 1.77 & SSGW03\\
    &            & 17.02.01 &  48  & 6.9 & V   & 2.7    & 2.8   &  2.0     & 3.8         & 0.36  & 1.48 & SSGW03\\ 
4.  &  1101+319 &  12.03.99 &  39  & 8.5 & NV  &        &       &          &             &       &      & Present work\\
    &   (RQQ)   &  04.04.00 &  22  & 5.6 & NV  &        &       &          &             &       &      & Present work\\
    &           &  21.04.01 &  21  & 6.1 & V   & 2.6    & 1.2   &  0.8     &             & 0.11  & 1.17 & SGSW03 \\
    &           &  22.04.01 &  21  & 5.8 & NV  &        &       &          &             &       &      & Present work\\
    & 1103$-$006&  17.03.99 &  23  & 3.8 & NV  &        &       &          &             &       &      & Present work\\
    &   (LDQ)   &  18.03.99 &  40  & 7.5 & V   & 3.1    & 2.4   &  0.6     &             & 0.55  & 1.62 & SGSW03 \\
    &           &  06.04.00 &  13  & 3.9 & PV  & 2.1    & 1.2   &          &             &       &      & SGSW03 \\
    &           &  25.03.01 &  28  & 7.2 & NV  &        &       &          &             &       &      & SGSW03 \\
    &           &  14.04.01 &  19  & 4.5 & NV  &        &       &          &             &       &      & Present work\\
    &           &  22.03.02 &  15  & 5.8 & PV  & 2.2    & 0.7   &          &             &       &      & SGSW03 \\ \hline
\end{tabular}
\end{table*}

\normalsize
\newpage

\begin{table*}
\noindent {\bf Table 2. {\it Continued}}
\vspace*{0.5cm}

\small
\begin{tabular}{llllcclcccccl} \hline
Set & Object     & Date     & N    & T & INOV  & C$_{eff}$ & $\psi$ & $\tau$ & P & $\eta_{obs}$ & $\delta_{obs}$ & Ref. to \\
No. &            &          &  &   (hr)  & Status&           &  \%    &  (hr)    & (hr) &      &           & DLCs   \\ \hline
4.  & 1216$-$010 & 11.03.02 &  22  & 8.0 & V   & 3.2    & 7.3   &  1.8     & 3.2         & 0.29  & 1.42 & SSGW03\\
    & (CDQ)      & 13.03.02 &  24  & 8.5 & V   & 2.6    & 3.8   &  1.2     & 2.2         & 0.16  & 1.26 & SSGW03\\
    &            & 15.03.02 &  11  & 3.9 & V   & 3.9    & 5.5   &  1.0     & 2.2         & 0.77  & 1.73 & SSGW03\\
    &            & 16.03.02 &  22  & 8.2 & V   & 6.6    & 14.1  &  $>$ 8.2 &             &       &      & SSGW03\\
    & 0735+178   & 26.12.98 &  49  & 7.8 & NV  &        &       &          &             &       &      & SGSW03 \\
    & (BL)       & 30.12.99 &  65  & 7.4 & NV  &        &       &          &             &       &      & Present work\\
    &            & 25.12.00 &  43  & 6.0 & NV  &        &       &          &             &       &      & Present work\\
    &            & 24.12.01 &  43  & 7.3 & V   & 2.8    & 1.0   &  $>$ 8.1 &             &  0.05 & 1.00 & SSGW03\\
5.  &  1029+329  & 13.03.99 &  45  & 8.4 & --- &        &       &          &             &       &      &       \\
    &   (RQQ)    & 02.03.00 &  19  & 5.0 & NV  &        &       &          &             &       &      & Present work    \\
    &            & 05.04.00 &  19  & 5.3 & V   & 4.3    & 1.3   & $>$6.2   &             & 0.13  & 1.21 & SGSW03    \\
    &            & 23.03.01 &  20  & 5.8 & NV  &        &       &          &             &       &      & Present work    \\
    &            & 06.03.02 &  31  & 8.5 & NV  &        &       &          &             &       &      & SGSW03   \\
    &            & 08.03.02 &  17  & 6.8 & V   & 2.8    & 1.1   &          &             &       &      & SGSW03  \\
    &  0709+370  & 20.01.01 &  29  & 6.5 & NV  &        &       &          &             &       &      & Present work   \\
    &   (LDQ)    & 21.01.01 &  29  & 6.2 & NV  &        &       &          &             &       &      & Present work   \\
    &            & 25.01.01 &  31  & 7.1 & NV  &        &       &          &             &       &      & Present work   \\
    &            & 20.12.01 &  49  & 7.9 & V   & 3.1    & 1.4   & 4.4      & 7.0         &  0.007&      & SGSW03  \\
    &            & 21.12.01 &  48  & 7.5 & NV  &        &       &          &             &       &      & Present work   \\
    & 0955+326   & 19.02.99 &  36  & 6.5 & NV  &        &       &          &             &       &      & Present work\\
    & (CDQ)      & 03.03.00 &  37  & 6.3 & NV  &        &       &          &             &       &      & Present work\\
    &            & 05.03.00 &  34  & 6.9 & PV  & 2.2    & 0.7   &          &             &       &      & SSGW03 \\
    & 0219+428   & 14.11.98 &  118 & 6.5 & V   & 6.0    & 5.4   &  $>$ 6.5 &             & 1.04  & 1.80 & SSGW03 \\
    & (BL)       & 13.11.99 &  123 & 5.7 & V   & $>$ 6.6& 5.5   &  $>$ 5.9 &             & 4.91  & 2.50 & SSGW03\\
    &            & 24.10.00 &  73  & 9.1 & V   & 5.8    & 4.3   &  $>$ 9.1 &             & 1.39  & 1.94 & SSGW03\\
    &            & 26.10.00 &  82  &10.1 & V   & 3.5    & 3.2   &      4.9 &             & 1.14  & 1.87 & SSGW03\\
    &            & 01.11.00 &  103 & 9.0 & V   & 2.9    & 2.2   &      3.9 &             & 0.26  & 1.39 & SSGW03\\
    &            & 24.11.00 &  71  & 5.1 & NV  &        &       &          &             &       &      & Present work\\
    &            & 01.12.00 &  59  & 5.1 & V   & $>$6.6 & 8.0   &  $>$ 5.1 &             & 2.14  & 2.12 & SSGW03\\
6.  &  0748+294  & 14.12.98 &  22  & 7.6 & NV  &        &       &          &             &       &      & Present work\\
    &  (RQQ)     & 13.01.99 &  56  & 8.3 & NV  &        &       &          &             &       &      & Present work\\
    &            & 09.12.99 &  26  & 5.1 & NV  &        &       &          &             &       &      & Present work\\
    &            & 24.11.00 &  28  & 5.4 & NV  &        &       &          &             &       &      & Present work\\
    &            & 01.12.00 &  32  & 6.0 & NV  &        &       &          &             &       &      & SGSW03 \\
    &            & 25.12.01 &  30  & 5.4 & NV  &        &       &          &             &       &      & SGSW03 \\
    & 0350$-$073 & 14.11.01 &  31  & 6.6 & NV  &        &       &          &             &       &      & Present work\\
    &  (LDQ)     & 15.11.01 &  26  & 5.5 & NV  &        &       &          &             &       &      & Present work\\
    &            & 18.11.01 &  25  & 5.7 & NV  &        &       &          &             &       &      & Present work\\
    & 1308+326   & 23.03.99 &  17  & 6.0 & --- &        &       &          &             &       &      &  \\
    &  (CDQ)     & 26.04.00 &  16  & 5.6 & NV  &        &       &          &             &       &      & Present work\\
    &            & 03.05.00 &  19  & 6.7 & --- &        &       &          &             &       &      &  \\
    &            & 17.03.02 &  19  & 7.7 & V   & 3.1    & 3.4   & 1.2,4.4  &             & 5.39  & 2.55 & SSGW03\\
    &            & 20.04.02 &  14  & 5.8 & NV  &        &       &          &             &       &      & Present work\\
    &            & 02.05.02 &  15  & 5.1 & NV  &        &       &          &             &       &      & Present work\\
    & 0235+164   & 13.11.98 &  36  & 4.4 & --- &        &       &          &             &       &      & \\
    &  (BL)      & 12.11.99 &  39  & 6.6 & V   & $>$6.6 & 12.8  &  3.6     &             & 2.48  & 2.18 & SSGW03\\
    &            & 14.11.99 &  34  & 6.2 & V   &  3.2   & 10.3  &  3.4     &             & 1.23  & 1.90 & SSGW03\\
    &            & 22.10.00 &  39  & 7.9 & V   &  2.6   & 7.6   &          &             & 1.71  & 2.03 & SSGW03\\
    &            & 28.10.00 &  29  & 6.8 & --- &        &       &          &             &       &      &  \\
7.  &  1017+279  & 14.03.99 &  43  & 7.3 & NV  &        &       &          &             &       &      & Present work\\
    &  (RQQ)     & 14.01.00 &  33  & 7.1 & NV  &        &       &          &             &       &      & Present work\\
    &            & 27.02.00 &  33  & 8.1 & NV  &        &       &          &             &       &      & Present work\\
    &  0012+305  & 18.01.01 &  17  & 3.6 & NV  &        &       &          &             &       &      & Present work\\
    &   (LDQ)    & 20.01.01 &  14  & 3.2 & NV  &        &       &          &             &       &      & Present work\\
    &            & 24.01.01 &  14  & 2.9 & NV  &        &       &          &             &       &      & Present work\\
    &            & 14.10.01 &  20  & 5.7 & NV  &        &       &          &             &       &      & Present work\\
    &            & 21.10.01 &  22  & 5.7 & NV  &        &       &          &             &       &      & SGSW03 \\
    &            & 22.10.01 &  24  & 6.2 & NV  &        &       &          &             &       &      & Present work\\
    & 1225+317   & 07.03.99 &  49  & 6.6 & NV  &        &       &          &             &       &      & Present work\\
    & (CDQ)      & 07.04.00 &  23  & 6.0 & NV  &        &       &          &             &       &      & Present work\\
    &            & 20.04.01 &  34  & 7.4 & NV  &        &       &          &             &       &      & Present work\\ \hline
\end{tabular}


\end{table*}

\newpage

\normalsize
\begin{table*}
\noindent {\bf Table 3. Results  for INOV and LTOV of the QSOs observed in this program}
\vspace*{0.5cm}

\begin{tabular}{lllrrrrrr} \hline
           &     &           &          &          &          &          &          &           \\
Object     &Type & Date      &  Q-S1    & Q-S2     &   Q-S3   & S1-S2    & S1-S3    & S2-S3     \\ 
           &     &           &  (mag)  & (mag)    & (mag)    & (mag)    & (mag)    & (mag)          \\ \hline
2349$-$014 & LDQ & 17.10.01 &    0.000 &    0.000 &    0.000 &    0.000 &    0.000 &    0.000  \\
           &     & 13.10.01 &    0.030 &    0.035 &    0.047 &    0.005 &    0.016 &    0.011  \\
           &     & 18.10.01 &    0.046 &    0.048 &    0.049 &    0.002 &    0.002 & $-$0.001  \\
1309+355   & LDQ & 01.04.01 &    0.000 &    0.000 &    0.000 &    0.000 &    0.000 &    0.000  \\
           &     & 02.04.01 &    0.004 &    0.003 &    0.000 &    0.000 & $-$0.004 & $-$0.004  \\
           &     & 25.03.99 &    0.107 &    0.097 &    0.128 & $-$0.010 &    0.021 &    0.031  \\
1215+303   & BL  & 25.04.01 &    0.000 &          &    0.000 &          &          &    0.000  \\
           &     & 19.04.01 &    0.108 &          &    0.098 &          &          & $-$0.010  \\
           &     & 31.03.00 &    0.201 &          &    0.206 &          &          &    0.006  \\
           &     & 25.02.00 &    0.344 &          &    0.322 &          &          & $-$0.022  \\
           &     & 20.03.99 &    0.522 &          &    0.468 &          &          & $-$0.054  \\
0514$-$005 & RQ  & 19.12.01 &    0.000 &    0.000 &    0.000 &    0.000 &    0.000 &    0.000  \\
           &     & 09.12.01 &    0.004 &    0.014 &    0.006 &    0.010 &    0.002 & $-$0.008  \\
           &     & 10.12.01 &    0.011 &    0.007 &    0.009 & $-$0.004 & $-$0.002 &    0.002  \\
1004+130   & LDQ & 18.02.01 &    0.000 &    0.000 &          &    0.000 &          &           \\
           &     & 24.03.01 & $-$0.003 &    0.001 &          &    0.004 &          &           \\
           &     & 29.03.00 &    0.002 &    0.005 &          &    0.003 &          &           \\
           &     & 30.03.00 &    0.021 &    0.028 &          &    0.007 &          &           \\
           &     & 16.03.99 &    0.077 &    0.089 &          &    0.012 &          &           \\
           &     & 27.02.99 &    0.081 &    0.099 &          &    0.018 &          &           \\
1128+310   &CDQ  & 18.01.01 &    0.000 &    0.000 &    0.000 &    0.000 &    0.000 &    0.000  \\
           &     & 09.03.02 &    0.135 &    0.136 &    0.130 &    0.001 & $-$0.005 & $-$0.006  \\
           &     & 10.03.02 &    0.037 &    0.136 &    0.132 & $-$0.001 & $-$0.006 & $-$0.004  \\
1252+020   &RQQ  & 09.03.00 &    0.000 &    0.000 &          &    0.000 &          &           \\
           &     & 03.04.00 &    0.006 &    0.002 &          & $-$0.003 &          &           \\
           &     & 18.03.02 &    0.087 &    0.082 &          & $-$0.005 &          &           \\
           &     & 26.04.01 &    0.177 &    0.183 &          &    0.006 &          &           \\
0134+329   & LDQ & 07.11.01 &    0.000 &    0.000 &    0.000 &    0.000 &    0.000 &    0.000  \\
           &     & 08.11.01 &    0.002 & $-$0.001 & $-$0.001 & $-$0.003 & $-$0.004 &    0.000  \\
           &     & 13.11.01 &    0.008 &    0.016 &    0.013 &    0.008 &    0.005 & $-$0.003  \\
1512+370   & LDQ & 21.04.02 &    0.000 &    0.000 &    0.000 &    0.000 &    0.000 &    0.000  \\
           &     & 23.04.02 &    0.031 &    0.025 &    0.023 & $-$0.005 & $-$0.008 & $-$0.002  \\
           &     & 01.05.02 &    0.033 &    0.027 &    0.030 & $-$0.006 & $-$0.003 &    0.003  \\
           &     & 27.03.02 &    0.054 &    0.048 &    0.047 & $-$0.005 & $-$0.007 & $-$0.002  \\
           &     & 23.03.02 &    0.085 &    0.082 &    0.082 & $-$0.003 & $-$0.003 &    0.001  \\
0851+202   & BL  & 31.12.99 &    0.000 &    0.000 &    0.000 &    0.000 &    0.000 &    0.000  \\
           &     & 29.12.98 &          &    0.601 &    0.695 &          &          &    0.094  \\
           &     & 28.03.00 &    0.936 &    0.928 &    0.906 & $-$0.009 & $-$0.031 & $-$0.020  \\
           &     & 17.02.01 &    1.631 &    1.628 &    1.624 & $-$0.004 & $-$0.007 & $-$0.003  \\
1101+319   & RQQ & 04.04.00 &          &    0.000 &    0.000 &          &          &    0.000  \\
           &     & 22.04.01 &          &    0.041 &    0.042 &          &          &    0.000  \\
           &     & 21.04.01 &          &    0.052 &    0.050 &          &          & $-$0.003  \\
           &     & 12.03.99 &          &    0.193 &    0.184 &          &          & $-$0.009  \\
1103$-$006 & LDQ & 06.04.00 &    0.000 &    0.000 &    0.000 &          &          &           \\
           &     & 17.03.99 &    0.018 &    0.018 & $-$0.004 &          &          &           \\
           &     & 18.03.99 &    0.021 &    0.021 & $-$0.001 &          &          &           \\
           &     & 14.04.01 &    0.316 &    0.314 & $-$0.003 &          &          &           \\
           &     & 25.03.01 &    0.318 &    0.314 & $-$0.005 &          &          &           \\
           &     & 22.03.02 &    0.329 &    0.324 & $-$0.006 &          &          &           \\
1216$-$010 & CDQ & 15.03.02 &    0.000 &    0.000 &          &    0.000 &          &           \\
           &     & 11.03.02 &    0.087 &    0.085 &          & $-$0.001 &          &           \\
           &     & 13.03.02 &    0.106 &    0.102 &          & $-$0.003 &          &           \\
           &     & 16.03.02 &    0.140 &    0.135 &          & $-$0.004 &          &           \\ \hline
\end{tabular}
\end{table*}

\begin{table*}
\noindent {\bf Table 3. {\it Continued}}
\vspace*{0.5cm}

\begin{tabular}{lllrrrrrr} \hline
           &     &           &          &          &          &          &          &            \\
Object     &Type & Date      &  Q-S1    & Q-S2     &   Q-S3   & S1-S2    & S1-S3    & S2-S3     \\ 
           &     &           &   (mag)  & (mag)    & (mag)    & (mag)    & (mag)    &  (mag)    \\ \hline
0735+178   & BL  & 26.12.98 &    0.000 &          &    0.000 &          &    0.000 &           \\
           &     & 30.12.99 &    0.457 &          &    0.465 &          &    0.008 &           \\
           &     & 24.12.01 &    0.720 &          &    0.715 &          & $-$0.005 &           \\
           &     & 25.12.00 &    1.145 &          &    1.147 &          &    0.002 &           \\
0709+370   & LDQ & 20.01.01 &    0.000 &    0.000 &    0.000 &    0.000 &    0.000 &    0.000  \\
           &     & 25.01.01 &    0.005 &    0.002 &    0.003 & $-$0.003 & $-$0.002 &    0.001  \\
           &     & 21.01.01 &    0.004 &    0.005 &    0.006 &    0.001 &    0.002 &    0.000  \\
           &     & 21.12.01 &    0.025 &    0.017 &    0.021 & $-$0.008 & $-$0.004 &    0.004  \\
           &     & 20.12.01 &    0.040 &    0.029 &    0.033 & $-$0.011 & $-$0.008 &    0.004  \\
0955+326   & CDQ & 19.02.99 &    0.000 &          &    0.000 &          &    0.000 &           \\
           &     & 03.03.00 &    0.044 &          &    0.057 &          &    0.013 &           \\
           &     & 05.03.00 &    0.047 &          &    0.060 &          &    0.013 &           \\
0219+428   & BL  & 24.10.00 &    0.000 &          &    0.000 &          &    0.000 &           \\
           &     & 26.10.00 &    0.255 &          &    0.256 &          &    0.001 &           \\
           &     & 24.11.00 &    0.277 &          &    0.281 &          &    0.003 &           \\
           &     & 01.11.00 &    0.331 &          &    0.330 &          & $-$0.001 &           \\
           &     & 13.11.99 &    0.449 &          &    0.449 &          &    0.000 &           \\
           &     & 01.12.00 &    0.622 &          &    0.631 &          &    0.010 &           \\
           &     & 14.11.98 &    0.691 &          &    0.684 &          & $-$0.007 &           \\
0748+294   & BL  & 13.01.99 &    0.000 &    0.000 &    0.000 &    0.000 &    0.000 &    0.000  \\
           &     & 25.12.01 &    0.231 &    0.221 &    0.214 & $-$0.010 & $-$0.017 & $-$0.007  \\
           &     & 24.11.00 &    0.231 &    0.221 &    0.210 & $-$0.009 & $-$0.021 & $-$0.012  \\
           &     & 01.12.00 &    0.237 &    0.226 &    0.209 & $-$0.010 & $-$0.028 & $-$0.017  \\
           &     & 14.12.98 &    0.239 &    0.236 &    0.207 & $-$0.002 & $-$0.031 & $-$0.029  \\
           &     & 09.12.99 &    0.242 &    0.238 &    0.234 & $-$0.004 & $-$0.008 & $-$0.004  \\
0350$-$073 & LDQ & 14.11.01 &    0.000 &    0.000 &    0.000 &    0.000 &    0.000 &    0.000  \\
           &     & 18.11.01 &    0.000 & $-$0.001 & $-$0.001 &    0.000 &    0.000 &    0.000  \\
           &     & 15.11.01 &    0.000 &    0.005 &    0.006 &    0.002 &    0.003 &    0.002  \\
1308+326   & CDQ & 26.04.00 &    0.000 &    0.000 &          &    0.000 &          &           \\
           &     & 03.05.00 &    0.016 &    0.009 &          & $-$0.006 &          &           \\
           &     & 23.03.99 &    0.204 &    0.203 &          &    0.000 &          &           \\ \hline
\end{tabular}
\end{table*}

\newpage

\begin{figure}

\noindent {\bf Figure 6.}  Differential light curves of radio lobe-dominated quasars. 
The labels are as in Fig.\ 5.
\end{figure}

\begin{figure}

\noindent {\bf Figure 6 {\it contiuned}} 
\end{figure}

\begin{figure}

\noindent {\bf Figure 6 {\it continued}} 
\end{figure}

\begin{figure}
\vspace*{-0.5cm}
\hspace*{-1.5cm}\psfig{file=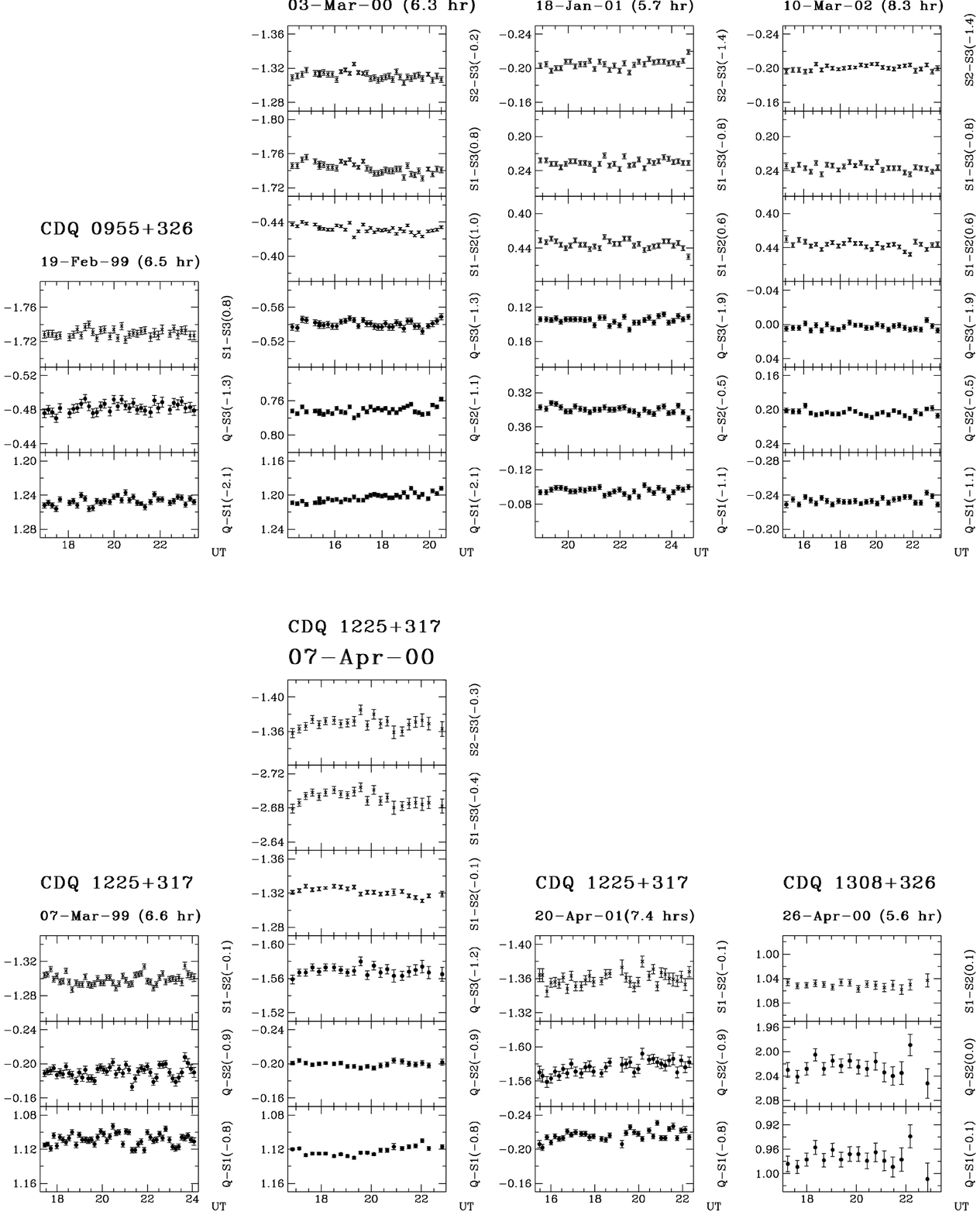}

\vspace*{-3.5cm}
\noindent {\bf Figure 7. } Differential light curves of radio core-dominated quasars.
The labels are as in Fig.\ 5.
\end{figure}

\begin{figure}
\vspace*{-0.5cm}
\hspace*{-1.5cm}\psfig{file=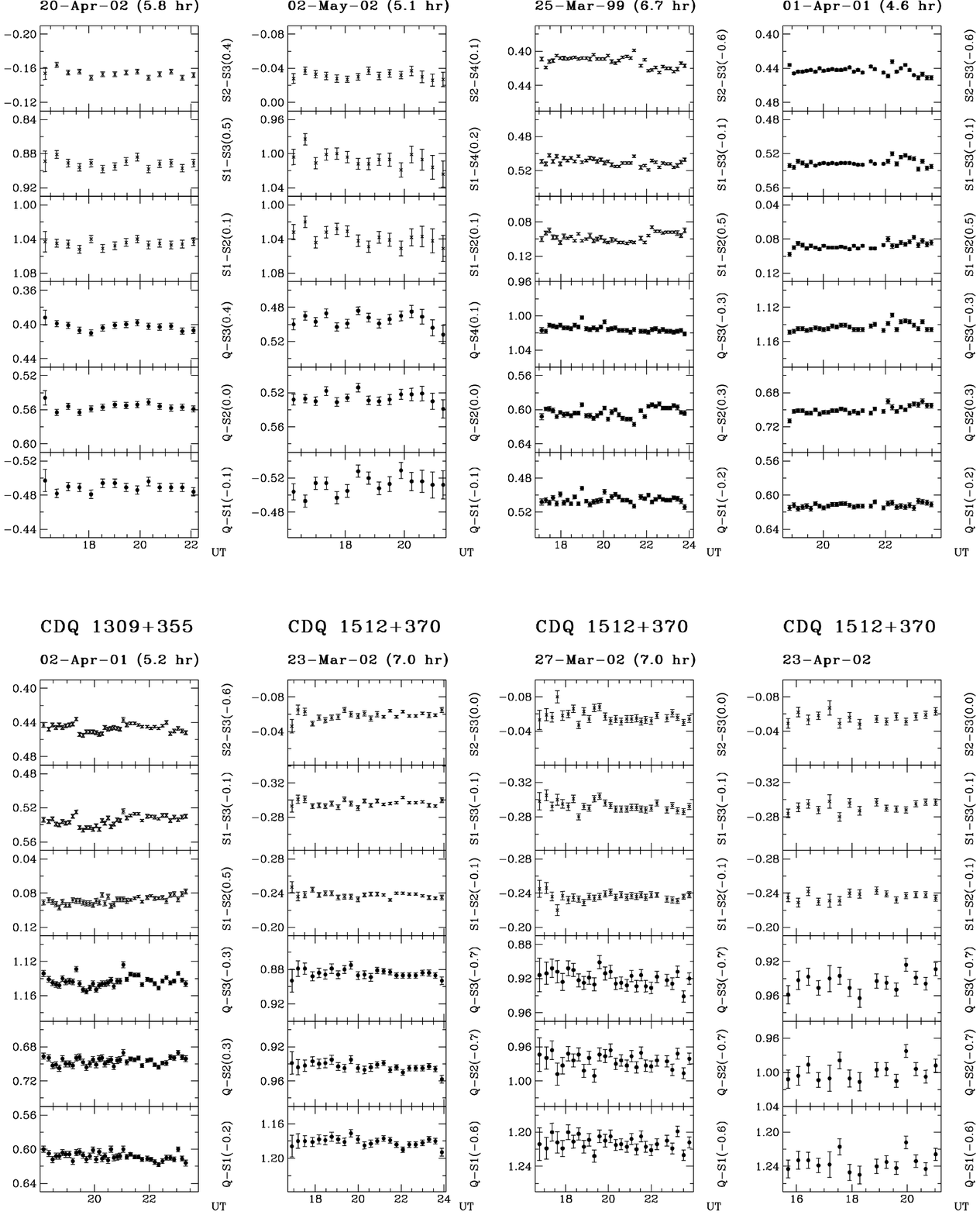}

\vspace*{-3.5cm}
\noindent {\bf Figure 7 {\it continued}}
\end{figure}

\begin{figure}
\vspace*{-0.5cm}
\hspace*{-1.5cm}\psfig{file=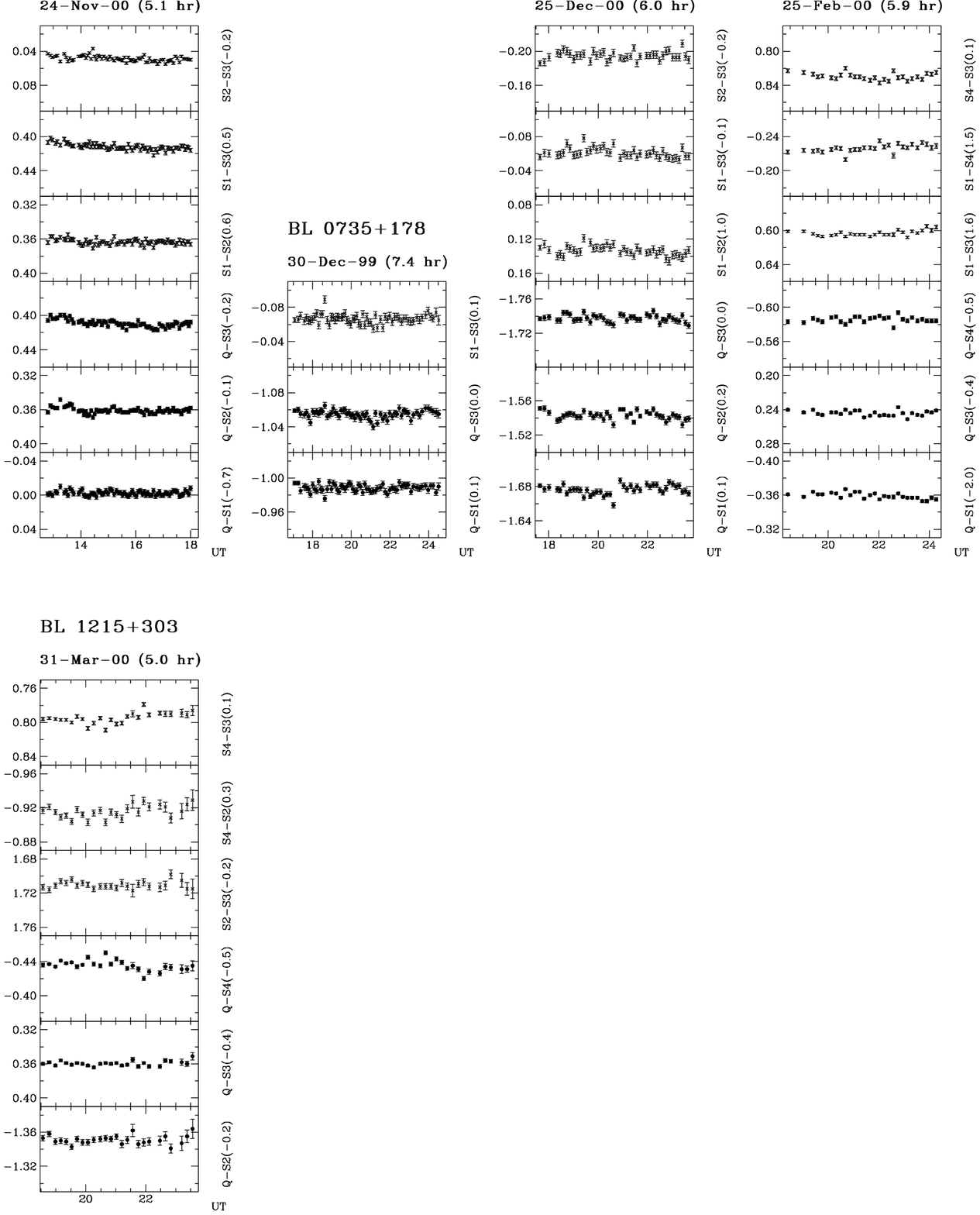}

\vspace*{-3.5cm}
\noindent {\bf  Figure 8.} Differential ligtcurves of BL lacertae objects. 
The labels are as in Fig.\ 5.
\end{figure}

\begin{figure}
\vspace*{-0.5cm}
\hspace*{-1.5cm}\psfig{file=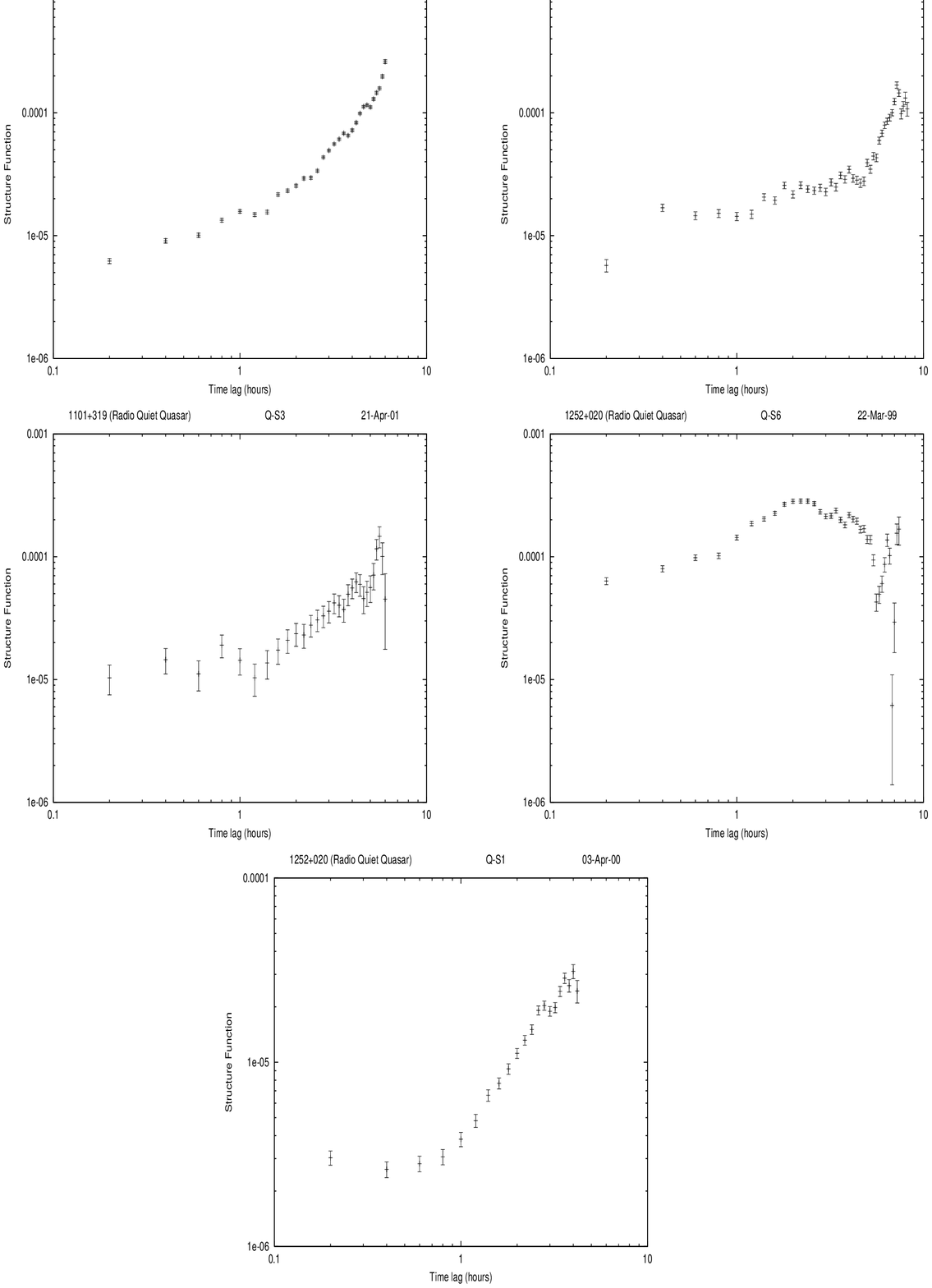}

\vspace*{-1.5cm}
\noindent {\bf Figure 9.} First order structure function of RQQs which have
shown INOV, given in increasing order of right ascension. 
\end{figure}

\begin{figure*}
\vspace*{-0.5cm}
\hspace*{-1.5cm}\psfig{file=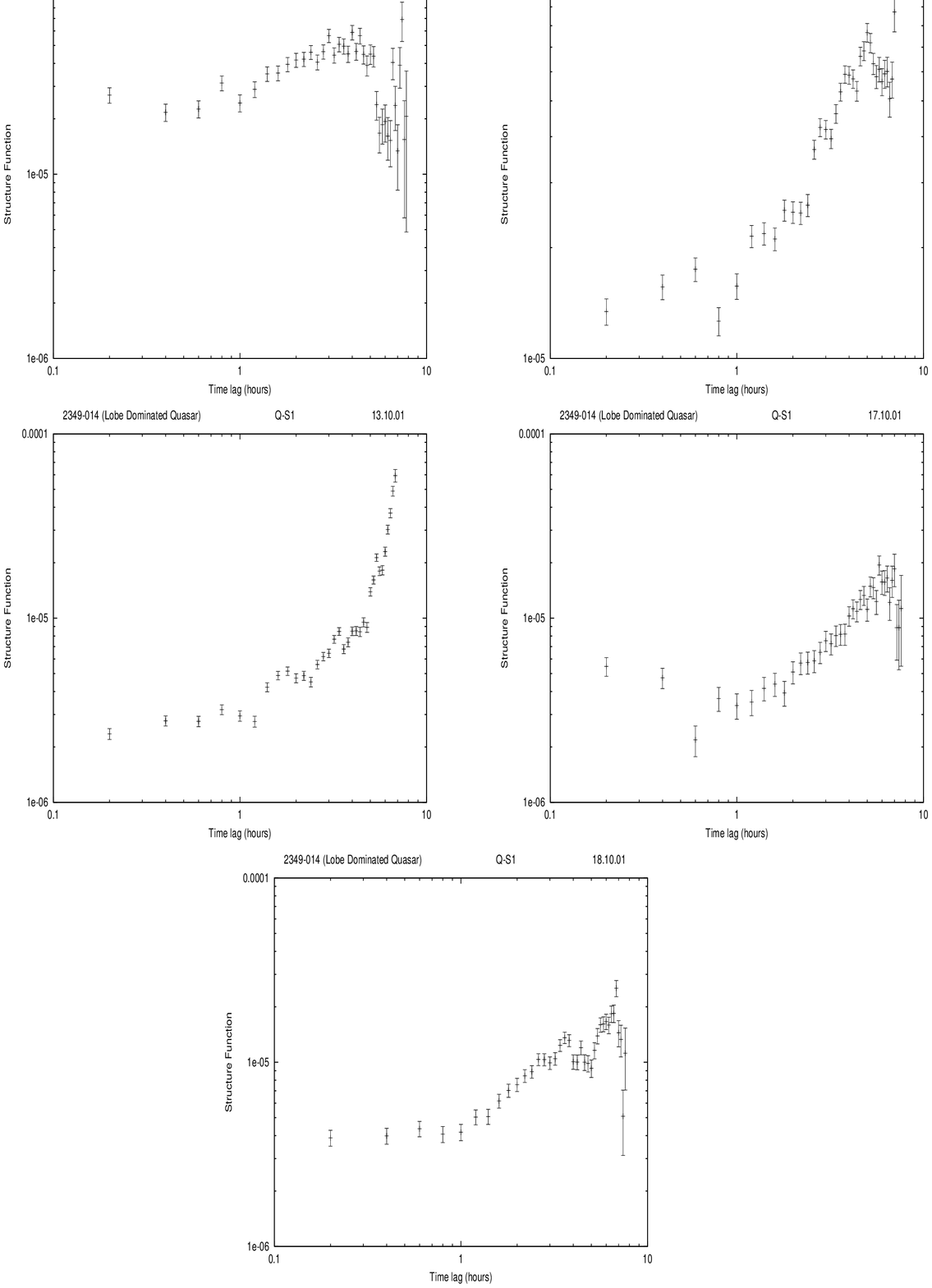}

\vspace*{-1.5cm}
\noindent {\bf Figure 10.} First order structure function of LDQs which have
shown INOV, given in increasing order of right ascension. 
\end{figure*}

\begin{figure}

\noindent {\bf Figure 11.} Long term variability of 1 core dominated quasar and 5 lobe dominated quasars 
observed in this programme.
\end{figure}

\begin{figure}

\noindent {\bf Figure 12.}  Long term variability of 6 radio-quiet quasars 
observed in this programme,
\end{figure}

\newpage

%

\begin{figure}
\vspace*{-0.5cm}
\hspace*{-1.5cm}\psfig{file=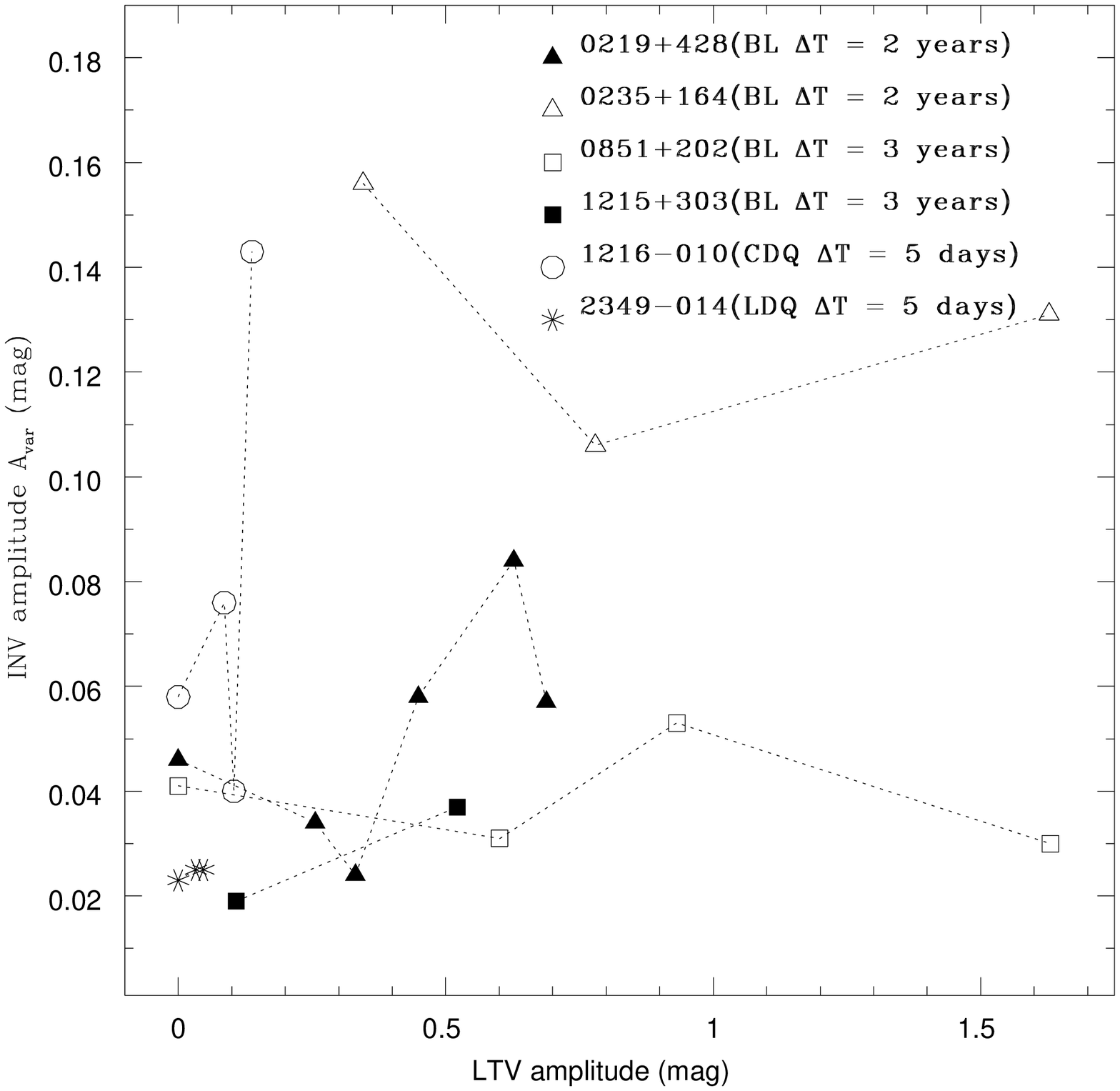}

\vspace*{0.2cm}
\noindent {\bf Figure 13.} Comparison of INOV against LTOV. 
\end{figure}

\begin{figure}
\vspace*{-0.5cm}
\hspace*{-1.5cm}\psfig{file=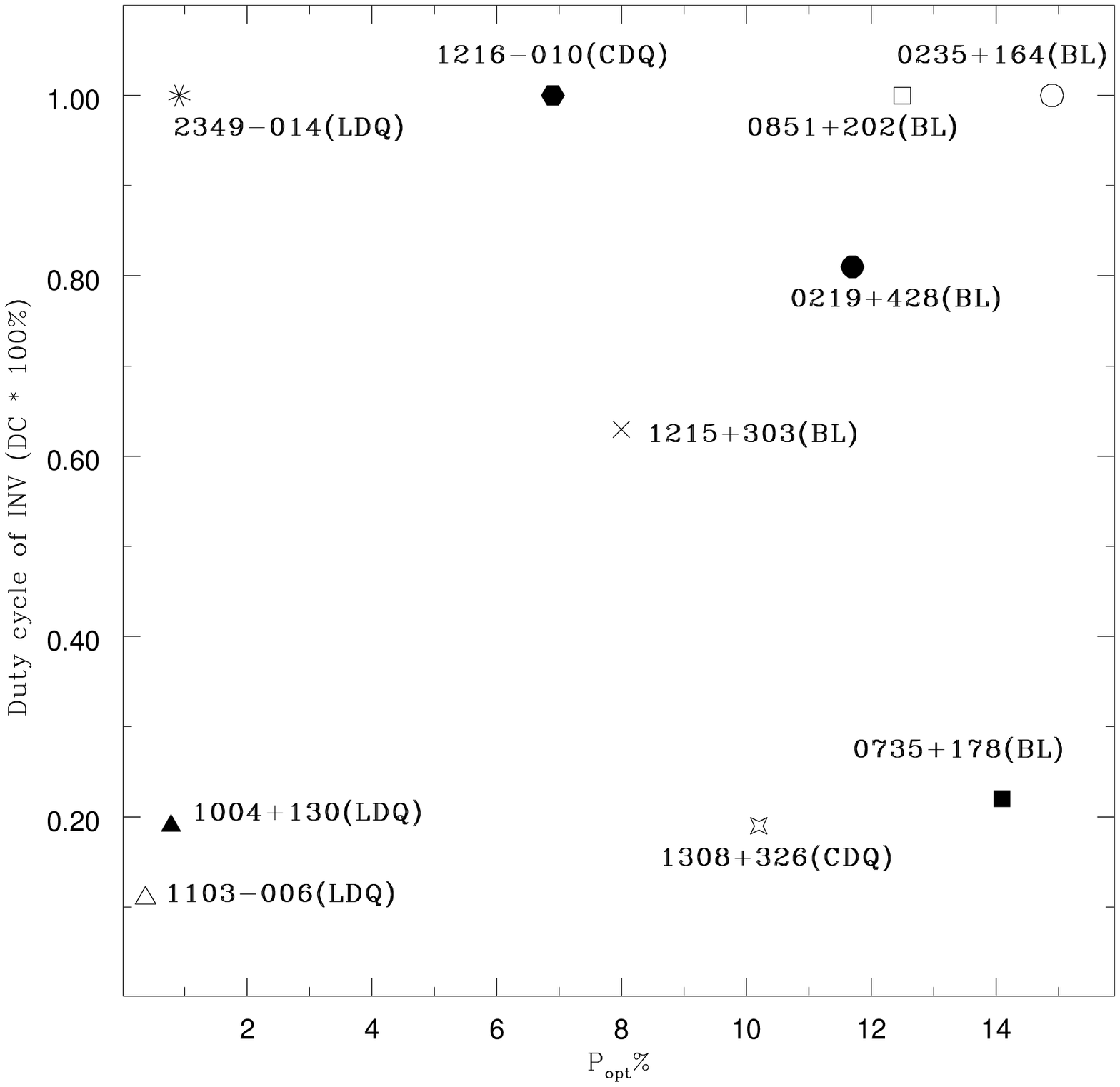}

\vspace*{-0.2cm}
\noindent {\bf Figure 14.} INOV duty cycles for objects which show 
INOV against their optical polarization.
\end{figure}

\end{document}